\documentclass{article}

\PassOptionsToPackage{numbers, compress}{natbib}

\usepackage[final]{neurips_2020}
\newcommand{\xhdr}[1]{{\noindent\bfseries #1}}

\usepackage[utf8]{inputenc} 
\usepackage[T1]{fontenc}    
\usepackage{hyperref}      
\usepackage{url}           
\usepackage{booktabs}      
\usepackage{amsfonts}      
\usepackage{nicefrac}     
\usepackage{adjustbox}
\usepackage{comment}

\usepackage{floatrow}
\newfloatcommand{capbtabbox}{table}[][\FBwidth]

\usepackage{graphicx}

\usepackage{subcaption}
\usepackage{booktabs}
\usepackage{url}
\usepackage{todonotes}
\usepackage{multirow}
\usepackage{tikz}

\usepackage{amsmath}
\usepackage{amssymb}
\usepackage{amsthm}
\usepackage{amsfonts}
\usepackage{thmtools}		
\usepackage{mleftright}
\usepackage{stmaryrd}
\usepackage{nicefrac}

\theoremstyle{definition}
\newtheorem{theorem}{Theorem}
\newtheorem{proposition}[theorem]{Proposition}
\newtheorem{lemma}[theorem]{Lemma}
\newtheorem{corollary}[theorem]{Corollary}

\usepackage{thm-restate}
\usepackage[mathic=true]{mathtools}
\usepackage{fixmath}
\usepackage{siunitx}
\usepackage{color}

\usepackage{pifont}
\newcommand{\cmark}{\ding{51}}%
\newcommand{\xmark}{\ding{55}}%

\usepackage{enumitem}
\setlist[enumerate]{noitemsep, topsep=0.5\topsep}
\setlist[description]{noitemsep, topsep=0.5\topsep}
\setlist[itemize]{noitemsep, topsep=0.5\topsep}

\usepackage{setspace}
\usepackage{ellipsis}
\usepackage{xspace}

\usepackage{algorithm}
\usepackage{algorithmicx}
\usepackage{algpseudocode}
\algdef{SE}{ParForAllLoop}{EndParForAllLoop}[1]{\textbf{parallel for}\(\mbox{#1}\) \textbf{do}}{\textbf{end}}

\usepackage{ifthen}
\newcommand{\CC}[1][]{$\text{C\hspace{-.25ex}}^{_{_{_{++}}}}
	\ifthenelse{\equal{#1}{}}{}{\text{\hspace{-.625ex}#1}}$}

\makeatletter
\def\thmt@refnamewithcomma #1#2#3,#4,#5\@nil{%
	\@xa\def\csname\thmt@envname #1utorefname\endcsname{#3}%
	\ifcsname #2refname\endcsname
	\csname #2refname\expandafter\endcsname\expandafter{\thmt@envname}{#3}{#4}%
	\fi
}
\makeatother
\usepackage[capitalise,noabbrev]{cleveref}   
\newcommand{\new}[1]{\emph{#1}}

\newcommand{\cO}{\ensuremath{{\mathcal O}}\xspace}

\newcommand{\bbR}{\ensuremath{\mathbb{R}}}

\newcommand{\bbN}{\ensuremath{\mathbb{N}}}

\newcommand{\RR}{\mathbb{R}}

\newcommand{\NN}{\mathbb{N}}

\newcommand{\ndelta}{\ensuremath{\overline{\delta}}}

\newcommand{\oms}{\{\!\!\{}
\newcommand{\cms}{\}\!\!\}}

\renewcommand{\vec}[1]{\mathbf{#1}}

\newcommand{\wl}{$1$-\textsf{WL}\xspace}
\newcommand{\kwl}{$k$-\textsf{WL}\xspace}

\newcommand{\deltakwl}{$\delta$-$k$-\textsf{WL}\xspace}
\newcommand{\kwlm}{k\textrm{-}\textsf{WL}\xspace}

\newcommand{\deltakwlm}{\delta\textrm{-}k\textrm{-}\textsf{WL}\xspace}
\newcommand{\localkwl}{$\delta$-$k$-\textsf{LWL}\xspace}
\newcommand{\pluskwl}{$\delta$-$k$-\textsf{LWL}$^+$\xspace}

\newcommand{\pluskwlm}{\delta\textrm{-}k\textrm{-}\textsf{LWL}^+\xspace}
\newcommand{\deltakwln}{$\delta$-$k$-\textsf{GNN}\xspace}
\newcommand{\localkwln}{$\delta$-$k$-\textsf{LGNN}\xspace}
\newcommand{\kwln}{$k$-\textsf{WL-GNN}\xspace}
\newcommand{\kgnn}{$k$\textrm{-}\textsf{GNN}\xspace}
\newcommand{\kign}{$k$\textrm{-}\textsf{IGN}\xspace}
\newcommand{\gnn}{\textsf{GNN}\xspace}
\newcommand{\mpnn}{\textsf{MPNN}\xspace}
\newcommand{\shp}{\textsf{SP}\xspace}
\newcommand{\gr}{\textsf{GR}\xspace}
\newcommand{\wloa}{\textsf{WLOA}\xspace}
\newcommand{\gin}{\textsf{GIN}\xspace}
\newcommand{\gine}{\textsf{GINE}\xspace}
\newcommand{\gineps}{\textsf{GIN-$\varepsilon$}\xspace}
\newcommand{\gineeps}{\textsf{GINE-$\varepsilon$}\xspace}
\newcommand{\UNR}{\textsf{UNR}\,}

\newtheorem{claim}[theorem]{Claim}
\newcommand{\deltaunr}{\delta\text{-}\UNR\xspace}
\newcommand{\localunr}{\textsf{L}\text{-}\UNR\xspace}
\newcommand{\plusunr}{\textsf{L}^{\!+}\text{-}\UNR\xspace}

\usetikzlibrary{hobby,arrows,decorations.pathmorphing,backgrounds,positioning,fit,petri,calc}
\usepackage[framemethod=tikz]{mdframed}

\definecolor{mycolor}{rgb}{0.122, 0.435, 0.698}

\newmdenv[innerlinewidth=0.5pt, roundcorner=4pt,linecolor=mycolor,innerleftmargin=6pt,
innerrightmargin=6pt,innertopmargin=6pt,innerbottommargin=6pt]{mybox}

\title{Weisfeiler and Leman go sparse: Towards scalable higher-order graph embeddings}

\author{Christopher Morris\thanks{CERC in Data Science for Real-Time Decision-Making, Polytechnique Montr{\'e}al} \And Gaurav Rattan\thanks{Department of Computer Science, RWTH Aachen University} \And Petra Mutzel\thanks{Department of Computer Science, University of Bonn}
}

\begin{document}

\maketitle

\begin{abstract}
Graph kernels based on the $1$-dimensional Weisfeiler-Leman algorithm and corresponding neural architectures recently emerged as powerful tools for (supervised) learning with graphs. However, due to the purely local nature of the algorithms, they might miss essential patterns in the given data and can only handle binary relations. The $k$-dimensional Weisfeiler-Leman algorithm addresses this by considering $k$-tuples, defined over the set of vertices, and defines a suitable notion of adjacency between these vertex tuples. Hence, it accounts for the higher-order interactions between vertices. However, it does not scale and may suffer from overfitting when used in a machine learning setting. Hence, it remains an important open problem to design WL-based graph learning methods that are simultaneously expressive, scalable, and non-overfitting. Here, we propose \new{local} variants and corresponding neural architectures, which consider a subset of the original neighborhood, making them more scalable, and less prone to overfitting. The expressive power of (one of) our algorithms is strictly higher than the original algorithm, in terms of ability to distinguish non-isomorphic graphs. Our experimental study confirms that the local algorithms, both kernel and neural architectures, lead to vastly reduced computation times, and prevent overfitting. The kernel version establishes a new state-of-the-art for graph classification on a wide range of benchmark datasets, while the neural version shows promising performance on large-scale molecular regression tasks.
\end{abstract}

\section{Introduction}
Graph-structured data is ubiquitous across application domains ranging from chemo- and bioinformatics~\cite{Barabasi2004,Sto+2020} to image~\cite{Sim+2017} and social network analysis~\cite{Eas+2010}. To develop successful machine learning models in these domains, we need techniques that can exploit the rich information inherent in the graph structure, as well as the feature information contained within nodes and edges. In recent years, numerous approaches have been proposed for machine learning with graphs---most notably, approaches based on \new{graph kernels}~\cite{Kri+2019} or using \new{graph neural networks} (GNNs)~\cite{Cha+2020,Gil+2017,Gro+2020}. Here, graph kernels based on the \new{$1$-dimensional Weisfeiler-Leman algorithm} (\wl)~\cite{Gro2017,Wei+1968}, and corresponding GNNs~\cite{Mor+2019,Xu+2018b} have recently advanced the state-of-the-art in supervised node and graph learning. Since the \wl operates via simple neighborhood aggregation, the purely local nature of these approaches can miss important patterns in the given data. Moreover, they are only applicable to binary structures, and therefore cannot deal with general $t$-ary structures, e.g., hypergraphs~\cite{Zho+2006} or subgraphs, in a straight-forward way. A provably more powerful algorithm (for graph isomorphism testing) is the \emph{$k$-dimensional Weisfeiler-Leman algorithm} (\kwl)~\cite{Cai+1992,Gro2017,Mar+2019}. The algorithm can capture more global, higher-order patterns by iteratively computing a coloring (or discrete labeling) for $k$-tuples, instead of single vertices, based on an appropriately defined notion of adjacency between two $k$-tuples. However, it fixes the cardinality of this neighborhood to $k\cdot n$, where $n$ denotes the number of vertices of a given graph. Hence, the running time of each iteration does not take the \emph{sparsity} of a given graph into account. Further, new neural architectures~\cite{Mar+2019,Mar+2019b} that possess the same power as the \kwl in terms of separating non-isomorphic graphs suffer from the same drawbacks, i.e., they have to resort to dense matrix multiplications. Moreover, when used in a machine learning setting with real-world graphs, the \kwl may capture the isomorphism type, which is the complete structural information inherent in a graph, after only a couple of iterations, which may lead to overfitting, see~\cite{Mor+2017}, and the experimental section of the present work.

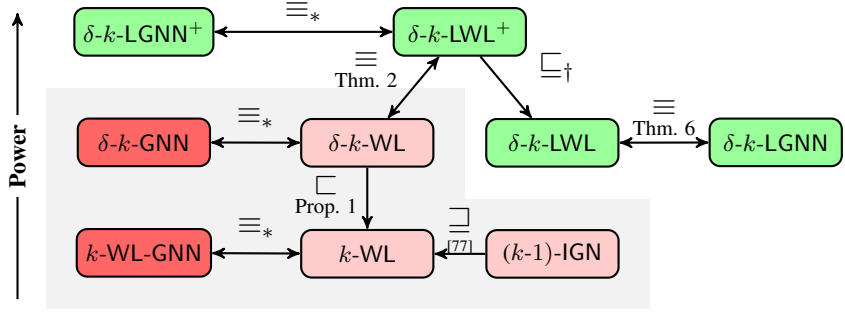
\begin{figure}[t]
	\begin{center}
		\tikzset{
			treenode/.style = {shape=rectangle, rounded corners,
				draw, align=center,
				minimum width=50pt,
			}
		}
		\trimbox{0pt 20pt 0pt 10pt}{
		\begin{tikzpicture}[scale=1.4,font=\footnotesize,  >=stealth', thick,sibling distance=15mm, level distance=30pt,minimum size=18pt,sibling distance=50pt]
		\node(a) [treenode,fill=green!40] {$\delta\text{-}k\text{-}\mathsf{LWL}^+$} 
		child { node(b) [treenode,fill=red!20] {$\delta\text{-}k\text{-}\mathsf{WL}$}
			child { node(d) [treenode,fill=red!20] {$k\text{-}\mathsf{WL}$}
				edge from parent node [left] {\large $\underset{\text{Prop. 1}}{\sqsubset}$} [->] }
			edge from parent node [left,->,label={[shift={(-0.3,-0.5)}]\large $\underset{\text{Thm. 2}}{\equiv}$}]{} [<->] }
		child { node(c) [treenode,fill=green!40] {$\delta$-$k\text{-}\mathsf{LWL}$}
		edge from parent node [left,->,label={[shift={(1.0,-0.4)}]\large $\sqsubseteq_{\dagger}$}]{} [->] };
		
		\node(e) at (-3.0,0) [treenode,fill=green!40] {$\delta\text{-}k\text{-}\mathsf{LGNN}^+$};
		\node(f) at (-3.0,-1.05) [treenode,fill=red!60] {$\delta\text{-}k\text{-}\mathsf{GNN}$};
		\node(g) at (-3.0,-2.1) [treenode,fill=red!60] {$k\text{-}\mathsf{WL\text{-}GNN}$};
		\node(h) at (.88,-2.11) [treenode,fill=red!20,grow=east] {($k$-$1)\text{-}\mathsf{IGN}$};
		\node(i) at (3.0,-1.05) [treenode,fill=green!40,grow=east] {$\delta$-$k\text{-}\mathsf{LGNN}$};
		{[shift={(1.0,0.3)}]Label}]
		
		\draw[<->] (i) -- (c) node[midway,label={[shift={(0.0,-.4)}]\large $\underset{\text{Thm. 6}}{\equiv}$}]{};
		\draw[<->] (a) -- (e) node[midway,label={[shift={(0.0,-.4)}]\large ${\equiv_{*}}$}]{};
		\draw[<->] (b) -- (f) node[midway,above]{\large $\equiv_{*}$};
		\draw[<->] (d) -- (g) node[midway,above]{\large $\equiv_{*}$};
		\draw[<-] (d) -- (h) node[midway,label={[shift={(0.0,-.5)}]\large $\underset{\text{\tiny\cite{Mar+2019}}}{\sqsupseteq}$}]{};
		
		\draw[->, thick] (-4.2,-2.54) -- (-4.2,0.18) node [midway,fill=white] {\rotatebox{90}{\textbf{Power}}};		
		
		\begin{pgfonlayer}{background}
		
		\node[fill=gray!10, inner sep=4.0mm,label=below:,fit=(b) (d) (g)] {};
		\node[fill=gray!10, inner sep=4.0mm,label=below:,fit=(d) (h)] {};

		\end{pgfonlayer}
		\end{tikzpicture}}
	\end{center}
	\caption{Overview of the power of proposed algorithms and neural architectures. The green and dark red nodes represent algorithms proposed in the present work. The grey region groups dense algorithms and neural architectures. \\
		$_*$---Follows directly from the proof of~\cref{equal}. $A \sqsubseteq B$ ($A \sqsubset B$, $A \equiv B$): algorithm $A$ is  more powerful (strictly more powerful, equally powerful) than $B$, 	$_{\dagger}$---Follows by definition, strictness open.}\label{overview}
\end{figure}

\xhdr{Present work} To address this, we propose a \new{local} version of the \kwl, the \new{local $\delta$-$k$-dimensional Weisfeiler-Leman algorithm} (\localkwl), which considers a subset of the original neighborhood in each iteration. The cardinality of the \new{local neighborhood} depends on the sparsity of the graph, i.e., the degrees of the vertices of a given $k$-tuple. We theoretically analyze the strength of a variant of our local algorithm and prove that it is strictly more powerful in distinguishing non-isomorphic graphs compared to the \kwl. Moreover, we devise a hierarchy of pairs of non-isomorphic graphs that a variant of the \localkwl can separate while the \kwl cannot. On the neural side, we devise a higher-order graph neural network architecture, the \localkwln, and show that it has the same expressive power as the \localkwl. Moreover, we connect it to recent advancements in learning theory for GNNs~\cite{Gar+2020}, which show that the \localkwl architecture has better generalization abilities compared to dense architectures based on the \kwl. See~\cref{overview} for an overview of the proposed algorithms.

Experimentally, we apply the discrete algorithms (or kernels) and the (local) neural architectures to supervised graph learning, and verify that both are several orders of magnitude faster than the global, discrete algorithms or dense, neural architectures, and prevent overfitting. The discrete algorithms establish a new state-of-the-art for graph classification on a wide range of small- and medium-scale classical datasets. The neural version shows promising performance on large-scale molecular regression tasks.

\xhdr{Related work}
In the following, we review related work from graph kernels and GNNs. We refer to~\cref{exprel} for an in-depth discussion of related work, as well as a discussion of theoretical results for the \kwl.

Historically, kernel methods---which implicitly or explicitly map graphs to elements of a Hilbert space---have been the dominant approach for supervised learning on graphs. Important early work in this area includes kernels based on random-walks~\cite{Gae+2003,Kas+2003,Kri+2017b}, shortest paths~\cite{Bor+2005}, and kernels based on the \wl~\cite{She+2011}. \citeauthor{Mor+2017}~\cite{Mor+2017} devised a local, set-based variant of the \kwl. However, the approach is (provably) weaker than the tuple-based algorithm, and they do not prove convergence to the original algorithm. For a thorough survey of graph kernels, see~\cite{Kri+2019}. Recently, graph neural networks (GNNs)~\cite{Gil+2017,Sca+2009} emerged as an alternative to graph kernels. Notable instances of this architecture include, e.g.,~\cite{Duv+2015,Ham+2017,Vel+2018}, and the spectral approaches proposed in, e.g.,~\cite{Bru+2014,Def+2015,Kip+2017,Mon+2017}---all of which descend from early work in~\cite{Kir+1995,Mer+2005,Spe+1997,Sca+2009}. A survey of recent advancements in GNN techniques can be found, e.g., in~\cite{Cha+2020,Wu+2019,Zho+2018}. Recently, connections to Weisfeiler-Leman type algorithms have been shown~\cite{Bar+2020,Che+2019,Gee+2020a,Gee+2020b,Mae+2019,Mar+2019,Mor+2019,Xu+2018b}. Specifically, the authors of~\cite{Mor+2019,Xu+2018b} showed that the expressive power of any possible GNN architecture is limited by the \wl in terms of distinguishing non-isomorphic graphs. \citeauthor{Mor+2019}~\cite{Mor+2019} introduced \emph{$k$-dimensional GNNs} (\kgnn) which rely on a message-passing scheme between subgraphs of cardinality $k$. Similar to~\cite{Mor+2017}, the paper employed a local, set-based (neural) variant of the \kwl, which is (provably) weaker than the variant considered here. Later, this was refined in~\cite{Mar+2019} by introducing \emph{$k$-order invariant graph networks} (\kign), based on \citeauthor{Mar+2019b}~\cite{Mar+2019b}, which are equivalent to the folklore variant of the \kwl~\cite{Gee+2020b,Gro2017} in terms of distinguishing non-isomorphic graphs. However, \kign may not scale since they rely on dense linear algebra routines. \citeauthor{Che+2019}~\cite{Che+2019} connect the theory of universal approximation of permutation-invariant functions and the graph isomorphism viewpoint and introduce a variation of the $2$-\textsf{WL}, which is more powerful than the former. Our comprehensive treatment of higher-order, sparse, (graph) neural networks for arbitrary $k$ subsumes all of the algorithms and neural architectures mentioned above.

\section{Preliminaries}

We briefly describe the Weisfeiler-Leman algorithm and, along the way, introduce our notation, see~\cref{prelim} for expanded preliminaries. As usual, let $[n] = \{ 1, \dotsc, n \} \subset \NN$ for $n \geq 1$, and let $\{\!\!\{ \dots\}\!\!\}$ denote a multiset. We also assume elementary definitions from graph theory (such as graphs, directed graphs, vertices, edges, neighbors,
trees, and so on). The vertex and the edge set of a graph $G$ are denoted by $V(G)$ and $E(G)$ respectively. The \new{neighborhood} of $v$ in $V(G)$ is denoted by $\delta(v) = N(v) = \{ u \in V(G) \mid (v, u) \in E(G) \}$. Moreover, its complement $\ndelta(v) = \{ u \in V(G) \mid (v, u) \notin E(G) \}$. We say that two graphs $G$ and $H$ are \new{isomorphic} ($G \simeq H$) if there exists an adjacency preserving bijection $\varphi \colon V(G) \to V(H)$, i.e., $(u,v)$ is in $E(G)$ if and only if 
$(\varphi(u),\varphi(v))$ is in $E(H)$, call $\varphi$ an \emph{isomorphism} from $G$ to $H$. If the graphs have vertex/edges labels, the isomorphism is additionally required to match these labels. A \new{rooted tree} is a tree with a designated vertex called \new{root} in which the edges are directed in such a way that they point away from the root. Let $p$ be a vertex in a directed tree then we call its out-neighbors \new{children} with parent $p$. Given a $k$-tuple of vertices $\vec{v} = (v_1,\dots,v_k)$, let $G[\vec{v}]$ denote the subgraph induced on the set $\{v_1,\dots,v_k\}$, where, the vertex $v_i$ is labeled with $i$, for $i$ in $[k]$.

\xhdr{Vertex refinement algorithms}\label{vr}
For a fixed positive integer $k$, let $V(G)^k$ denote the set of $k$-tuples of vertices of $G$. A \new{coloring} of $V(G)^k$ is a mapping $C \colon V(G)^k \to \mathbb{N}$, i.e., we assign a number (or color) to every tuple in $V(G)^k$. The \new{initial coloring} $C_{0}$ of $V(G)^k$ is specified by the isomorphism types of the tuples, i.e., two tuples $\vec{v}$ and $\vec{w}$ in $V(G)^k$ get a common color iff the mapping $v_i \to w_i$ induces an isomorphism between the labeled subgraphs $G[\vec{v}]$ and $G[\vec{w}]$. A \new{color class} corresponding to a color $c$ is the set of all tuples colored $c$, i.e., the set $C^{-1}(c)$. For $j$ in $[k]$, let $\phi_j(\vec{v},w)$ be the $k$-tuple obtained by replacing the $j^{\textrm{th}}\!$ component of $\vec{v}$ with the vertex $w$. That is, $\phi_j(\vec{v},w) = (v_1, \dots, v_{j-1}, w, v_{j+1}, \dots, v_k)$. If $\vec{w} = \phi_j(\vec{v},w)$ for some $w$ in $V(G)$, call $\vec{w}$ a $j$-\new{neighbor} of $\vec{v}$ (and vice-versa). The \emph{neighborhood} of $\vec{v}$ is then defined as the set of all tuples $\vec{w}$ such that $\vec{w} = \phi_j(\vec{v},w)$ for some $j$ in $[k]$ and $w$ in $V(G)$. The \emph{refinement} of a coloring $C \colon V(G)^k \to \mathbb{N}$, denoted by $\widehat{C}$, is a coloring $\widehat{C} \colon V(G)^k \to \mathbb{N}$ defined as follows. 
For each $j$ in $[k]$, collect the colors of the $j$-neighbors of $\vec{v}$ as a multiset $S_j=\{\!\! \{  C(\phi_j(\vec{v},w)) \mid w \in V(G) \} \!\!\}$.
Then, for a tuple $\vec{v}$, define $\widehat{C}(\vec{v}) = (C(\vec{v}), M(\vec{v}))$, where $M(\vec{v})$ is the $k$-tuple $(S_1,\dots,S_k)$. For consistency, the strings $\widehat{C}(\vec{v})$ thus obtained are lexicographically sorted and renamed as integers. Observe that the new color $\widehat{C}(\vec{v})$ of $\vec{v}$ is solely dictated by the color histogram of its neighborhood and the previous color of $\vec{v}$. In general, a different mapping $M(\cdot)$ could be used, depending on the neighborhood information that we would like to aggregate. 

\xhdr{The $\boldsymbol{k}$-dim. Weisfeiler-Leman} For $k\geq 2$, the \kwl computes a coloring $C_\infty \colon V(G)^k \to \mathbb{N}$ of a given graph $G$, as follows.\footnote{We define the $1$-\textsf{WL} in the next subsection.} To begin with, the initial coloring $C_0$ is computed. Then, starting with $C_0$, successive refinements $C_{i+1} = \widehat{C_i}$ are computed until convergence. That is,
\[
C_{i+1}(\vec{v}) = (C_i(\vec{v}), M_i(\vec{v})),
\]
where 
\begin{equation}\label{mi}
M_i(\vec{v}) =   \big( \{\!\! \{  C_i(\phi_1(\vec{v},w)) \mid w \in V(G) \} \!\!\}, \dots, \{\!\! \{  C_i(\phi_k(\vec{v},w)) \mid w \in V(G) \} \!\!\} \big).
\end{equation}
The successive refinement steps are also called \new{rounds} or \new{iterations}. Since the disjoint union of the color classes form a partition of $V(G)^k$, there must exist a finite $\ell \leq |V(G)|^k$ such that $C_{\ell} = \widehat{C_{\ell}}$. In the end, the \kwl outputs $C_\ell$ as the \emph{stable coloring} $C_\infty$. The \kwl \new{distinguishes} two graphs $G$ and $H$ if, upon running the \kwl on their disjoint union $G \,\dot\cup\, H$, there exists a color $c$ in $\mathbb{N}$ in the stable coloring such that the corresponding color class $S_c$ satisfies
$|V(G)^k \cap S_c| \neq |V(H)^k \cap S_c|$, i.e., there exist an unequal number of $c$-colored tuples in $V(G)^k$ and $V(H)^k$. Hence, two graphs distinguished by the \kwl must be non-isomorphic. See \cref{wl_app} for its relation to the folklore \kwl.

\xhdr{The $\boldsymbol{\delta}$-$\boldsymbol{k}$-dim. Weisfeiler-Leman} Let $\vec{w} = \phi_j(\vec{v},w)$ be a $j$-{neighbor} of $\vec{v}$. Call $\vec{v}$ a \new{local} $j$-neighbor of $\vec{w}$ if $w$ is adjacent to the replaced vertex $v_j$. Otherwise, call $\vec{v}$ a \new{global} $j$-neighbor of $\vec{w}$. For tuples $\vec{v}$ and $\vec{w}$ in $V(G)^k$, let the function $\textrm{adj}((\vec{v},\vec{w}))$ evaluate to $\textsc{L}$ or $\textsc{G}$, depending on whether $\vec{w}$ is a local or a global neighbor, respectively, of $\vec{v}$. The $\delta$-$k$-dimensional \new{Weisfeiler-Leman algorithm}, denoted by \deltakwl, is a variant of the classic \kwl which \emph{differentiates} between the local and the global neighbors during neighborhood aggregation \cite{Mal2014}. Formally, the \deltakwl algorithm refines a coloring $C_i$ (obtained after $i$ rounds) via the aggregation function 
\begin{equation}\label{mid}
\begin{split}
M^{\delta,\ndelta}_i(\vec{v}) =   \big( &\{\!\! \{  (C_i({\phi_1(\vec{v},w)}, \textrm{adj}(\vec{v}, \phi_1(\vec{v},w)     )) \mid w \in V(G) \} \!\!\}, \dots, \\ &\{\!\! \{  (C_i({\phi_k(\vec{v},w)}, \textrm{adj}(\vec{v},  \phi_k(\vec{v},w))) \hspace{-0.55pt}\mid w \in V(G) \}  \!\!\} \big),
\end{split}	
\end{equation}
instead of the \kwl aggregation specified by~\cref{mi}. We define the $1$-\textsf{WL} to be the $\delta$-1-\textsf{WL}, which is commonly known as color refinement or naive vertex classification.

\xhdr{Comparison of $\boldsymbol{k}$-\textsf{WL} variants} Let $A_1$ and $A_2$ denote two vertex refinement algorithms, we write $A_1 \sqsubseteq A_2$ if $A_1$ distinguishes between all non-isomorphic pairs $A_2$ does, and $A_1 \equiv A_2$ if both directions hold. The corresponding strict relation is denoted by $\sqsubset$. The following result shows that the \deltakwl is strictly more powerful than the \kwl for $k \geq 2$ (see~\cref{proofrel1} for the proof).  
\begin{proposition}\label{rel1} 
For $k \geq 2 $, the following holds: 	
\begin{equation*}
\deltakwlm \sqsubset \kwlm.
\end{equation*}
\end{proposition}
\section{Local $\boldsymbol{\delta}$-$\boldsymbol{k}$-dimensional Weisfeiler-Leman algorithm}\label{lwl}

In this section, we define the new \new{local $\delta$-$k$-dimensional Weisfeiler-Leman algorithm} (\localkwl). This variant of the \deltakwl considers only local neighbors during the neighborhood aggregation process, and discards any information about the global neighbors. Formally, the \localkwl algorithm refines a coloring $C_i$ (obtained after $i$ rounds) via the aggregation function, 
\begin{equation}\label{eqnmidd}
\begin{split}
M^{\delta}_i(\vec{v}) =   \big( \{\!\! \{ C_{i}(\phi_1(\vec{v},w)) \mid w \in N(v_1) \} \!\!\}, \dots, \{\!\! \{  C_{i}(\phi_k(\vec{v},w)) \mid w \in N(v_k) \}  \!\!\} \big),
\end{split}
\end{equation}		
instead of~\cref{mid}, hence considering only the local $j$-neighbors of the tuple $\vec{v}$ in each iteration. The indicator function $\mathrm{adj}$ used in~\cref{mid} is trivially equal to $\textsc{L}$ here, and is thus omitted. The coloring function for the \localkwl is then defined by
\begin{equation*}\label{ck}
C^{k,\delta}_{i+1}(\vec{v}) = (C^{k,\delta}_{i}(\vec{v}), M^{\delta}_i(\vec{v})).
\end{equation*} 

We also define the \pluskwl, a minor variation of the \localkwl. Later, we will show that the \pluskwl is equivalent in power to the \deltakwl (\cref{loco}). 
Formally, the \pluskwl algorithm refines a coloring $C_i$ (obtained after $i$ rounds) via the aggregation function, 
\begin{equation}\label{middp}
\begin{split}
M^{\delta,+}(\vec{v}) =   \big( &\{\!\! \{ (C_i(\phi_1(\vec{v},w)), \#_{i}^1(\vec{v},\phi_1(\vec{v},w))) \hspace{0.4pt}\mid w \in N(v_1) \} \!\!\}, \dots, \\ &\{\!\! \{  (C_i(\phi_k(\vec{v},w)), \#_{i}^k(\vec{v},\phi_k(\vec{v},w))) \mid w \in N(v_k) \}  \!\!\} \big),
\end{split}
\end{equation}
instead of \localkwl aggregation defined in~\cref{eqnmidd}. 
Here, the function
\begin{equation}\label{sharp}
\#_{i}^j(\vec{v}, \vec{x}) = \big|\{ \vec{w} \colon \, \vec{w} \sim_j \vec{v}, \, C_{i}(\vec{w}) = C_{i}(\vec{x})   \} \big|,
\end{equation}
where $\vec{w} \sim_j \vec{v}$ denotes that $\vec{w}$ is $j$-neighbor of $\vec{v}$, for $j$ in $[k]$. Essentially, $\#_{i}^j(\vec{v}, \vec{x})$ counts the number of $j$-neighbors (local or global) of $\vec{v}$ which have the same color as $\vec{x}$ under the coloring $C_i$ (i.e., after $i$ rounds). For a fixed $\vec{v}$,
the function $\#_{i}^j(\vec{v},\cdot)$ is uniform over the set $S \cap N_j$, where $S$ is a color class obtained after $i$ iterations of the \pluskwl and $N_j$ denotes the set of $j$-neighbors of $\vec{v}$. 
Note that after the stable partition has been reached $\#_{i}^j(\vec{v})$ will not change anymore. Intuitively, this variant captures local and to some extent global information, while still taking the sparsity of the underlying graph into account. Moreover, observe that each iteration of the \pluskwl has the same asymptotic running time as an iteration of the \localkwl, and that the information of the $\#$ function is already implicitly contained in~\cref{mid}.

The following theorem shows that the local variant \pluskwl is at least as powerful as the \deltakwl when restricted to the class of connected graphs. The possibly slower convergence leads to advantages in a machine learning setting, see~\cref{neural,exp}, and also~\cref{prac_app} for a discussion of practicality, running times, and remaining challenges.
\begin{theorem}\label{loco}
	For the class of connected graphs, the following holds for all $k \geq 1$:
	\begin{equation*}
	\pluskwlm \equiv \deltakwlm.
	\end{equation*}
\end{theorem}	
Along with~\cref{rel1}, this establishes the superiority of the \pluskwl over the \kwl. 
\begin{corollary}\label{cloco} For the class of connected graphs, the following holds for all $k \geq 2$:	\begin{equation*}
	\pluskwlm \sqsubset \kwlm.
	\end{equation*}
\end{corollary}
In fact, the proof of \cref{rel1} shows that the infinite family of graphs $G_k, H_k$ witnessing the strictness condition can even be distinguished by the \localkwl,
for each corresponding $k \geq 2$. We note here that the restriction to connected graphs can easily be circumvented by adding a specially marked vertex, which is connected to every other vertex in the graph.

\xhdr{Kernels based on vertex refinement algorithms}
After running the \localkwl (and the other vertex refinements algorithms), the concatenation of the histogram of colors in each iteration can be used as a feature vector in a kernel computation. 
Specifically, in the histogram for every color $c$ in $\bbN$ there is an entry containing the number of nodes or $k$-tuples that are colored with $c$.

\xhdr{Local converges to global: proof of~\cref{loco}} The main technique behind the proof is to construct tree-representations of the colors assigned by the \kwl (or its variants). Given a graph $G$, a tuple $\vec{v}$, and an integer $\ell \geq 0$, the \emph{unrolling tree} of the graph $G$ \new{at $\vec{v}$ of depth $\ell$} is a rooted directed tree $\UNR[G,\vec{s},\ell]$ (with vertex and edge labels) which encodes the color assigned by \kwl to the tuple $\vec{v}$ after $\ell$ rounds, see~\cref{locodes} for a formal definition and~\cref{rolling_app} for an illustration. The usefulness of these tree representations is established by the following lemma. Formally, let $\vec{s}$ and $\vec{t}$ be two $k$-vertex-tuples in $V(G)^k$.
\begin{lemma}\label{encwl}
	The colors of $\vec{s}$ and $\vec{t}$ after $\ell$ rounds of \kwl are identical if and only if the unrolling tree $\UNR[G,\vec{s},\ell]$ is isomorphic to the unrolling tree $\UNR[G,\vec{t},\ell]$.  
\end{lemma}
For different \kwl variants, the construction of these unrollings are slightly different, since an unrolling tree needs to faithfully represent the corresponding aggregation process for generating new colors. For the variants \deltakwl, \localkwl, and \pluskwl, we define respective unrolling trees $\deltaunr[G,\vec{s},\ell]$, $\localunr[G,\vec{s},\ell]$, and $\plusunr[G,\vec{s},\ell]$ along with analogous lemmas, as above, stating their correctness/usefulness. Finally, we show that for \emph{connected graphs}, the $\deltaunr$ unrolling trees (of sufficiently large depth) at two tuples $\vec{s}$ and $\vec{t}$ are identical only if the respective \pluskwl unrolling trees (of sufficiently larger depth) are identical, as shown in the following lemma. 
\begin{lemma}\label{pluseqtodelta}
	Let $G$ be a connected graph, and let $\vec{s}$ and $\vec{t}$ in $V(G)^k$. If the stable colorings of $\vec{s}$ and $\vec{t}$ under \pluskwl are identical, 
	then the stable colorings of $\vec{s}$ and $\vec{t}$ under \deltakwl are also identical. 
\end{lemma}
Hence, the local algorithm \pluskwl is at least as powerful as the global \deltakwl, for connected graphs, i.e., $\pluskwlm \sqsubseteq \deltakwlm$. 
The exact details and parameters of this proof can be found in the Appendix.

\section{Higher-order neural architectures}\label{neural}
Although the discrete kernels defined in the previous section are quite powerful, they are limited due to their fixed feature construction scheme, hence suffering from poor adaption to the learning task at hand and the inability to handle continuous node and edge labels in a meaningful way. Moreover, they often result in high-dimensional embeddings forcing one to resort to non-scalable, kernelized optimization procedures. 
This motivates our definition of a new neural architecture, called \new{local $\delta$-$k$-}\textsf{GNN} (\localkwln).
Given a labeled graph $G$, let each tuple $\vec{v}$ in $V(G)^k$ be annotated with an initial feature $f^{(0)}(\vec{v})$ determined by its isomorphism type. 
In each layer $t > 0$,  we compute a new feature $f^{(t)}(\vec{v})$ as 
\begin{equation*}\label{gnngeneral}
f^{W_1}_{\text{mrg}}\Big(f^{(t-1)}(\vec{v}) ,f^{W_2}_{\text{agg}}\big(\oms f^{(t-1)}(\phi_1(\vec{v},w)) \mid w \in \delta(v_1)\cms, \dots, \oms f^{(t-1)}(\phi_k(\vec{v},w)) \mid w \in \delta(v_k) \cms \big) \!\Big),
\end{equation*}
in  $\bbR^{1 \times e}$ for a tuple $\vec{v}$, where $W_1^{(t)}$ and $W_2^{(t)}$ are learnable parameter matrices from $\bbR^{d \times e}$.\footnote{For clarity of presentation we omit biases.}. Moreover, $f^{W_2}_{\text{mrg}}$ and the permutation-invariant $f^{W_1}_{\text{agg}}$ can be arbitrary (permutation-invariant) differentiable functions, responsible for merging and aggregating the relevant feature information, respectively. Initially, we set $f^{(0)}(\vec{v})$ to a one-hot encoding of the (labeled) isomorphism type of $G[\vec{v}]$. Note that we can naturally handle discrete node and edge labels as well as directed graphs, see~\cref{contfeat} on how to deal with continuous information. The following result demonstrates the expressive power of the \deltakwln, in terms of distinguishing non-isomorphic graphs. 
\begin{theorem}\label{equal}
	Let $(G, l)$ be a labeled graph. Then for all \mbox{$t\geq 0$} there exists a sequence of weights $\mathbf{W}^{(t)}$ such that 
	\begin{equation*}
	C^{k,\delta}_{t}(\vec{v}) = C^{k,\delta}_{t}(\vec{w}) \iff f^{(t)}(\vec{v}) = f^{(t)}(\vec{w}).
	\end{equation*}
	Hence, for all graphs, the following holds for all $k \geq 1$:
	\begin{equation*}
	\text{\localkwln} \equiv \text{\localkwl}.
	\end{equation*}
\end{theorem}     
Moreover, the \deltakwln inherits the main strength of the \localkwl, i.e., it can be implemented using sparse matrix multiplication. Note that it is not possible to come up with an architecture, i.e., instantiations of $f^{W_1}_{\text{mrg}}$ and  $f^{W_2}_{\text{agg}}$, such that it becomes more powerful than the \localkwl, see~\cite{Mor+2019}. However, all results from the previous section can be lifted to the neural setting. That is, one can derive neural architectures based on the \pluskwl, \deltakwl, and \kwl, called \localkwln\hspace{-3pt}$^+$, \deltakwln, and \kwln, respectively, and prove results analogous to~\cref{equal}.

\xhdr{Incorporating continous information}\label{contfeat}
Since many real-world graphs, e.g., molecules, have continuous features (real-valued vectors) attached to vertices and edges, using a one-hot encoding of the (labeled) isomorphism type is not a sensible choice. Let $a \colon V(G) \rightarrow \RR^{1\times d}$ be a function such that each vertex $v$ is annotated with a feature $a(v)$ in $\bbR^{1\times d}$, and let $\vec{v} = (v_1, \dots, v_k)$ be a $k$-tuple of vertices. Then we can compute an inital feature
\begin{equation}\label{encode}
f^{(0)}(\vec{v}) = f^{W_3}_{\text{enc}}\big( (a(v_1), \dots, a(v_k)) \big),
\end{equation}
for the tuple $\vec{v}$. Here, $f_{\text{enc}} \colon \big(\RR^{1\times d}\big)^k \to \RR^{1\times e}$ is an arbitrary differentiable, parameterized function, e.g., a multi-layer perceptron or a standard GNN aggregation function, that computes a joint representation of the $k$ node features $a(v_1), \dots, a(v_k)$. Moreover, it is also straightforward to incorporate the labeled isomorphism type and continuous edge label information. We further explore this in the experimental section.

\xhdr{Generalization abilities of the  neural architecture}
\citeauthor{Gar+2020}~\cite{Gar+2020}, studied the generalization abilities of a standard GNN architecture for binary classification using a margin loss. Under mild conditions, they bounded the empirical Rademacher complexity as
$\tilde{\cO}\! \left( \nicefrac{rdL}{\sqrt{m}\gamma} \right)\!,$
where $d$ is the maximum degree of the employed graphs, $r$ is the number of components of the node features, $L$ is the number of layers, and $\gamma$ is a parameter of the loss function. It is straightforward to transfer the above bound to the higher-order (local) layer from above. Hence, this shows that local, sparsity-aware, higher-order variants, e.g., \localkwln, exhibit a smaller generalization error compared to dense, global variants like the \kwln.

\section{Practicality, barriers ahead, and possible road maps}\label{prac_app}

As~\cref{loco} shows, the \pluskwl and its corresponding neural architecture, the \localkwln$\!\!^+$, have the same power in distinguishing non-isomorphic graphs as \deltakwl. Although for dense graphs, the local algorithms will have the same running time, for sparse graphs, the running time for each iteration can be upper-bounded by $|n^k|\cdot kd$, where $d$ denotes the maximum or average degree of the graph. Hence, the local algorithm takes the sparsity of the underlying graph into account, resulting in improved computation times compared to the non-local \deltakwl and the \kwl (for the same number of iterations). These observations also translate into practice, see~\cref{exp}. The same arguments can be used in favor of the \localkwl and \localkwln, which lead to even sparser algorithms. 

\xhdr{Obstacles} The biggest obstacle in applying the algorithms to truly large graphs is the fact that the algorithm considers all possible $k$-tuples leading to a lower bound on the running time of $\Omega(n^k)$. Lifting the results to the folklore \kwl, e.g.,~\cite{Mar+2019}, only ``shaves off one dimension''. Moreover, applying higher-order algorithms for large $k$ might lead to overfitting issues, see also~\cref{exp}. 

\xhdr{Possible solutions} Recent sampling-based approaches for graph kernels or GNNs, see, e.g.,~\cite{Che+2018,Che+2018a,Ham+2017,Hua+2018,Mor+2017} address the dependence on $n^k$, while appropriate pooling methods along the lines of~\cref{encode} address the overfitting issue. Finally, new directions from the theory community, e.g.,~\cite{Gro+2020a} paint further directions, which might result in more scalable algorithms.

\section{Experimental evaluation}\label{exp}

Our intention here is to investigate the benefits of the local, sparse algorithms, both kernel and neural architectures, compared to the global, dense algorithms, and standard kernel and GNN baselines.
More precisely, we address the following questions:\\
\xhdr{Q1} Do the local algorithms, both kernel and neural architectures, lead to improved classification and regression scores on real-world benchmark datasets compared to global, dense algorithms and standard baselines? \\
\xhdr{Q2}  Does the \pluskwl lead to improved classification accuracies compared to the  \localkwl? Does it lead to higher computation times?\\
\xhdr{Q3}  Do the local algorithms prevent overfitting to the training set?\\
\xhdr{Q4} How much do the local algorithms speed up the computation time compared to the non-local algorithms or dense neural architectures?

The source code of all methods and evaluation procedures is available at \url{https://www.github.com/chrsmrrs/sparsewl}. 

\xhdr{Datasets} To evaluate kernels, we use the following, well-known, small-scale datasets: \textsc{Enzymes}~\cite{Sch+2004,Bor+2005}, \textsc{IMDB-Binary}, \textsc{IMDB-Multi}~\cite{Yan+2015a}, \textsc{NCI1}, \textsc{NCI109}~\cite{Wal+2008}, \textsc{PTC\_FM}~\cite{Hel+2001}, \textsc{Proteins}~\cite{Dob+2003,Bor+2005}, and \textsc{Reddit-Binary}~\cite{Yan+2015a}. To show that our kernels also scale to larger datasets, we additionally used the mid-scale datasets: \textsc{Yeast}, \textsc{YeastH}, \textsc{UACC257}, \textsc{UACC257H}, \textsc{OVCAR-8}, \textsc{OVCAR-8H}~\cite{Yan+2008}. For the neural architectures, we used the large-scale molecular regression datasets \textsc{Zinc}~\cite{Dwi+2020,Jin+2018a} and \textsc{Alchemy}~\cite{Che+2020}.
To further compare to the (hierarchical) \kgnn~\cite{Mor+2019} and \kign~\cite{Mar+2019}, and show the benefits of our architecture in presence of continuous features, we used the \textsc{QM9}~\cite{Ram+2014,Wu+2018} regression dataset.\footnote{We opted for comparing on the \textsc{QM9} dataset to ensure a fair comparison concerning hyperparameter selection.}  All datasets can be obtained from~\url{http://www.graphlearning.io}~\cite{Mor+2020}. See~\cref{datasets} for further details.

\xhdr{Kernels} We implemented the \localkwl, \pluskwl, \deltakwl, and  \kwl kernel for $k$ in $\{2,3\}$. We compare our kernels to the Weisfeiler-Leman subtree kernel (\wl)~\cite{She+2011}, the Weisfeiler-Leman Optimal Assignment kernel (\wloa)~\cite{Kri+2016}, the graphlet kernel (\gr)~\cite{She+2009}, and the shortest-path kernel~\cite{Bor+2005} (\shp). All kernels were (re-)implemented in \CC[11]. For the graphlet kernel, we counted (labeled) connected subgraphs of size three. We followed the evaluation guidelines outlined in~\cite{Mor+2020}. We also provide precomputed Gram matrices for easier reproducability.

\xhdr{Neural architectures} We used the \gin and \gineps architecture~\cite{Xu+2018b} as neural baselines. For data with (continuous) edge features, we used a $2$-layer MLP to map them to the same number of components as the node features and combined them using summation (\gine and \gineeps). For the evaluation of the neural architectures of~\cref{neural}, \localkwln, \deltakwln, and \kwln, we implemented them using \textsc{PyTorch Geometric}~\cite{Fey+2019}, using a  Python-wrapped \CC[11] preprocessing routine to compute the computational graphs for the higher-order GNNs.\footnote{We opted for not  implementing the \localkwln$\!\!^+$ as it would involve precomputing $\#$.} We used the \gineps layer to express $f^{W_1}_{\text{mrg}}$ and $f^{W_2}_{\text{aggr}}$ of~\cref{gnngeneral}. 

See~\cref{protocol} for a detailed description of all evaluation protocols and hyperparameter selection routines. 
\begin{table}[t]\centering			
	\caption{Classification accuracies in percent and standard deviations,  \textsc{OOT}--- Computation did not finish within one day, \textsc{OOM}--- Out of memory.}
	\label{t2}	
	\resizebox{1.0\textwidth}{!}{ 	\renewcommand{\arraystretch}{1.05}
		\begin{tabular}{@{}c <{\enspace}@{}lcccccccc@{}}	\toprule
			& \multirow{3}{*}{\vspace*{4pt}\textbf{Method}}&\multicolumn{8}{c}{\textbf{Dataset}}\\\cmidrule{3-10}
			& & {\textsc{Enzymes}}         &  {\textsc{IMDB-Binary}}      & {\textsc{IMDB-Multi}}           & {\textsc{NCI1}}       & {\textsc{NCI109}}           & 
			{\textsc{PTC\_FM}}         & {\textsc{Proteins}}         &
			{\textsc{Reddit-Binary}  } \\	\toprule
			\multirow{4}{*}{\rotatebox{90}{\hspace*{-3pt}Baseline}} & \gr     &  29.7   \scriptsize	$\pm 0.6$  & 58.9	   \scriptsize 	$ 	\pm 1.0 $  &    39.0   \scriptsize	$	\pm 0.8$ & 66.1  \scriptsize	$\pm 0.4$     & 66.3    \scriptsize	 $	\pm 0.2$     &    61.3      \scriptsize	$	\pm	1.1$   &    71.2      \scriptsize	$	\pm 0.6	$     &  60.0    \scriptsize	$	\pm	0.2$     \\ 
			& \shp  & 40.7  \scriptsize	$	\pm 0.9	$   &  58.5   \scriptsize	$	\pm	0.4$  &  39.4   \scriptsize	$	\pm 0.3	$     & 74.0   \scriptsize	$	\pm 0.3	$     & 73.0  \scriptsize	$ 	\pm 0.4	$     & 61.3  \scriptsize	$	\pm 1.3	$ & 75.6   \scriptsize	$\pm 0.5	$ & 84.6 \scriptsize   	$\pm 0.3	$ \\ 
			& \textsf{$1$-WL}         & 50.7   \scriptsize $\pm 1.2$      & 72.5   \scriptsize	 $	\pm	0.5$  &  50.0      \scriptsize 	 $	\pm	0.5$    & 84.2  \scriptsize	 $	\pm 0.3	$ & 84.3   \scriptsize	 $\pm 0.3 	$ & \bf{62.6} \scriptsize	 $ 	\pm 2.0 	$  &  72.6 \scriptsize	$	\pm 1.2 $      & 72.8   \scriptsize	 $	\pm 0.5	$     \\ 						
			& \textsf{WLOA}            &    56.8    \scriptsize $\pm 1.6 $      &   72.7   \scriptsize	 $	\pm 0.9$  &   50.1       \scriptsize	 $	\pm 0.7	$    &   84.9      \scriptsize	 $	\pm 0.3	$ &  85.2    \scriptsize	 $\pm 0.3	$ &  61.8   \scriptsize	 $	\pm 1.5	$  &     73.2       \scriptsize	$	\pm 0.6	$      &    88.1     \scriptsize	 $	\pm	0.4$     \\ 		
			\cmidrule{2-10}		
			\multirow{2}{*}{\rotatebox{90}{\hspace*{-3pt}Neural}} & \textsf{Gin-$0$} & 38.8  \scriptsize $\pm 1.7$ &    72.7  \scriptsize $\pm 0.9$ & 49.9 \scriptsize $\pm 0.8$ & 78.5 \scriptsize $\pm 0.5$ & 76.7  \scriptsize $\pm 0.8$ & 58.2   \scriptsize $\pm 3.3$ &  71.3 \scriptsize $\pm 0.9$ & 89.8  \scriptsize $\pm 0.6$
			\\ 
			& \textsf{Gin-$\varepsilon$} & 39.4 \scriptsize $\pm 1.7$ &   72.9        \scriptsize	$\pm 0.6 $  & 49.6    \scriptsize $\pm 0.9$& 78.6  \scriptsize $\pm 0.3$ & 77.0  \scriptsize $\pm 0.5$ &  57.7   \scriptsize $\pm 2.0$ &  71.1   \scriptsize  $\pm 0.8$& 90.3  \scriptsize $\pm 0.3$ \\ 
			\cmidrule{2-10}	
			\multirow{4}{*}{\rotatebox{90}{Global}} 	&
			\textsf{$2$-WL}        &   36.7   \scriptsize  $\pm 1.7$ &  68.2  \scriptsize  $\pm 1.1$&  48.1  \scriptsize    $\pm 0.5$&       67.1 \scriptsize  $\pm 0.3$ &  67.5  \scriptsize  $ \pm 0.2$ & 62.3 \scriptsize  $\pm 1.6$ &  75.0   \scriptsize  $\pm 0.8$ & \textsc{Oom} \\
			&  \textsf{$3$-WL}        	&    42.3 \scriptsize $\pm 1.1$& 67.8 \scriptsize $\pm 0.8$ & 47.0   \scriptsize $\pm 0.7$ &\textsc{Oot} &\textsc{Oot} & 61.5    \scriptsize $\pm 1.7$ &\textsc{Oom} & \textsc{Oom} \\            
			\cmidrule{2-10}			
			& \textsf{$\delta$-$2$-WL}  	& 37.5  \scriptsize $\pm 1.2$   &  68.1   \scriptsize $\pm 1.1$&   47.9     \scriptsize  $\pm 0.7$ &      67.0    \scriptsize  $\pm 0.5$ &  67.2   \scriptsize  $\pm 0.4$&   61.9  \scriptsize  $\pm 0.9$& 75.0 \scriptsize  $\pm 0.4$ & \textsc{OOM} \\  
			& \textsf{$\delta$-$3$-WL}          		& 43.0  \scriptsize $\pm 1.4$& 67.5 \scriptsize $\pm 1.0$ &  47.3 \scriptsize $\pm 0/9$ &\textsc{Oot} &\textsc{Oot} &   61.2  \scriptsize $\pm 2.0$ &\textsc{Oom} & \textsc{Oom} \\                                                   
			\cmidrule{2-10}		
			\multirow{4}{*}{\rotatebox{90}{Local}}    		
			& \textsf{$\delta$-$2$-LWL}         	& 56.6    \scriptsize  $\pm 1.2$&  73.3 \scriptsize  $\pm 0.5$ & 50.2 \scriptsize  $\pm 0.6 $& 84.7 \scriptsize  $\pm 0.3$&   84.2   \scriptsize  $\pm 0.4$ & 60.3   \scriptsize  $\pm 3.2$&   75.1  \scriptsize  $\pm 0.3 $& 89.7 \scriptsize        $\pm 0.4$  \\      		
			& \textsf{$\delta$-$2$-LWL$^+$}         	&  52.9  \scriptsize  $\pm 1.4$&  75.7 \scriptsize  $\pm 0.7 $&  62.5   \scriptsize  $\pm 1.0 $&   \bf{91.4}   \scriptsize  $\pm 0.2$ &  \bf{89.3}        \scriptsize  $\pm 0.2$&  \bf{62.6}  \scriptsize  $\pm 1.6$&   \textbf{79.3}    \scriptsize $\pm 1.1 $&   \bf{91.1}    \scriptsize  $\pm 0.5 $\\     
			& \textsf{$\delta$-$3$-LWL} 	&  \bf{57.6}   \scriptsize  $\pm 1.2 $& 72.8 \scriptsize  $\pm  1.2  $&   49.3     \scriptsize  $\pm 1.0 $ &  83.4   \scriptsize  $\pm 0.2  $&  82.4   \scriptsize  $\pm 0.4$ &    61.3    \scriptsize  $\pm 1.6$& \textsc{Oom} & \textsc{Oom} \\ 	
			& \textsf{$\delta$-$3$-LWL$^+$}         	& 56.8   \scriptsize  $\pm 1.2$&  \bf{76.2}     \scriptsize  $\pm  0.8$& \bf{64.2}       \scriptsize  $\pm 0.9 $ &  82.7   \scriptsize  $\pm $0.5 &  81.9   \scriptsize  $\pm 0.4$& 61.3 \scriptsize  $\pm 2.0$& \textsc{Oom}& \textsc{Oom} \\    
			\bottomrule
	\end{tabular}}
\end{table}	
	
\begin{table}
	\begin{center}
		\begin{subfigure}[c,top]{0.48\textwidth}	\centering
			\label{t2n}	
\caption{\label{t3_short} Training versus test accuracy of local and global kernels for a subset of the datasets.
}
\resizebox{0.95\textwidth}{!}{ 	\renewcommand{\arraystretch}{1.05}
\begin{tabular}{@{}c <{\enspace}@{}lccc@{}}	\toprule
	& \multirow{3}{*}{\vspace*{4pt}\textbf{Set}}&\multicolumn{3}{c}{\textbf{Dataset}}\\\cmidrule{3-5}
	& & {\textsc{Enzymes}}         &  {\textsc{IMDB-Binary}}      & {\textsc{IMDB-Multi}}  \\	\toprule
	\multirow{2}{*}{\rotatebox{25}{\textsf{$\delta$-2-WL}}}    		
	& Train & 91.2  & 83.8  &57.6  \\ 
	
	& Test & 37.5   &  68.1 &   47.9    \\              
	\cmidrule{2-5}
	\multirow{2}{*}{\rotatebox{25}{\textsf{$\delta$-$2$-LWL}}}    		
	& Train  & 98.8  &  83.5 & 59.9   \\ 
	& Test  &  56.6  &  73.3  & 50.2   \\   
	\cmidrule{2-5}
	\multirow{2}{*}{\rotatebox{25}{\textsf{$\delta$-$2$-LWL$^+$}}}    		
	& Train  & 99.5  & 95.1 &  86.5   \\ 
	& Test   	&  52.9 &  75.7 &  62.5   \\       
	\bottomrule
\end{tabular}
}
		\end{subfigure}\hspace{10pt}
		\begin{subfigure}[c,top]{0.48\textwidth}	\centering
							\caption{Mean MAE (mean std. MAE, logMAE) on large-scale (multi-target) molecular regression tasks.\label{neural_short_tt}}
						\resizebox{.85\textwidth}{!}{\renewcommand{\arraystretch}{1.05}
				\begin{tabular}{@{}c <{\enspace}@{}lcc@{}}	\toprule
					& \multirow{3}{*}{\vspace*{4pt}\textbf{Method}}&\multicolumn{2}{c}{\textbf{Dataset}}\\\cmidrule{3-4}
					&    &  {\textsc{Zinc (Full)}}     & {\textsc{alchemy (Full)}}       \\	\toprule
					\multirow{4}{*}{\rotatebox{90}{Baseline\hspace{3pt}}}
					& \gineeps  & 0.084                                                                                                                                                                                                                                     
					\scriptsize $\pm 0.004$                                                                                                                                                                                                                                  & 0.103 {\scriptsize $\pm 0.001$} -2.956 {\scriptsize $\pm 0.029$} \\	
					\cmidrule{2-4}
					&  \textsf{$2$-WL-GNN}    & 0.133 \scriptsize $\pm 0.013$    & {0.093                                                                                                                                                                                                                                       
						\scriptsize $\pm 0.001$}  {-3.394                                                                                                                                                                                                                                    
						\scriptsize $\pm 0.035$}                                                                                                                                \\
					&  $\delta$-$2$-\textsf{GNN}      &\textbf{0.042} {\scriptsize $\pm 0.003$}   & \textbf{0.080}                                                                           {\scriptsize $\pm 0.001$} -3.516 {\scriptsize $\pm 0.021$} \\			
					\cmidrule{2-4}	
					\multirow{2}{*}{\rotatebox{90}{}}   & $\delta$-$2$-\textsf{LGNN} &  0.045  \scriptsize $\pm 0.006$ & 0.083 {\scriptsize $\pm 0.001$} -3.476  {\scriptsize $\pm 0.025$} \\
					\bottomrule
			\end{tabular}}\vspace{14pt}
		\end{subfigure}
	\end{center}
		\caption{Additional results for kernel and neural approaches.}\label{add_tt}
\end{table}

\xhdr{Results and discussion} In the following we answer questions \textbf {Q1} to \textbf{Q4}.

\begin{figure}[t]
	\begin{center}
		\begin{subfigure}[c]{0.25\textwidth}\centering
			\vspace{2pt}
			\resizebox{.95\textwidth}{!}{\renewcommand{\arraystretch}{1.05}
				\begin{tabular}{@{}c <{\enspace}@{}lc@{}}	\toprule
					&\textbf{Method}&  \textsc{QM9} \\	\toprule
					\multirow{6}{*}{\rotatebox{90}{Baseline}}
					& \gineeps   &  0.081 {\scriptsize $\pm 0.003$} \\	
					& \mpnn & 0.034 {\scriptsize $\pm 0.001$}   \\	
					& \textsf{$1$-$2$-GNN} &  0.068   {\scriptsize $\pm 0.001  $}  \\   
					& \textsf{$1$-$3$-GNN} &  0.088 {\scriptsize $\pm 0.007$}  \\
					& \textsf{$1$-$2$-$3$-GNN} & 0.062  {\scriptsize $\pm 0.001$} \\	
					& \textsf{$3$-IGN} &  0.046 {\scriptsize $\pm 0.001$}  \\	
					\midrule
					& \textsf{$\delta$-$2$-LGNN} & \textbf{0.029} {\scriptsize $\pm 0.001$} \\	
					\bottomrule
			\end{tabular}}	\vspace{6pt}
					\caption{Mean std.~MAE compared to~\cite{Gil+2017,Mar+2019,Mor+2019}.}
			\label{t2nn_short}
		\end{subfigure}
		\begin{subfigure}[c]{0.32\textwidth}
			\centering
			\includegraphics[scale=0.28]{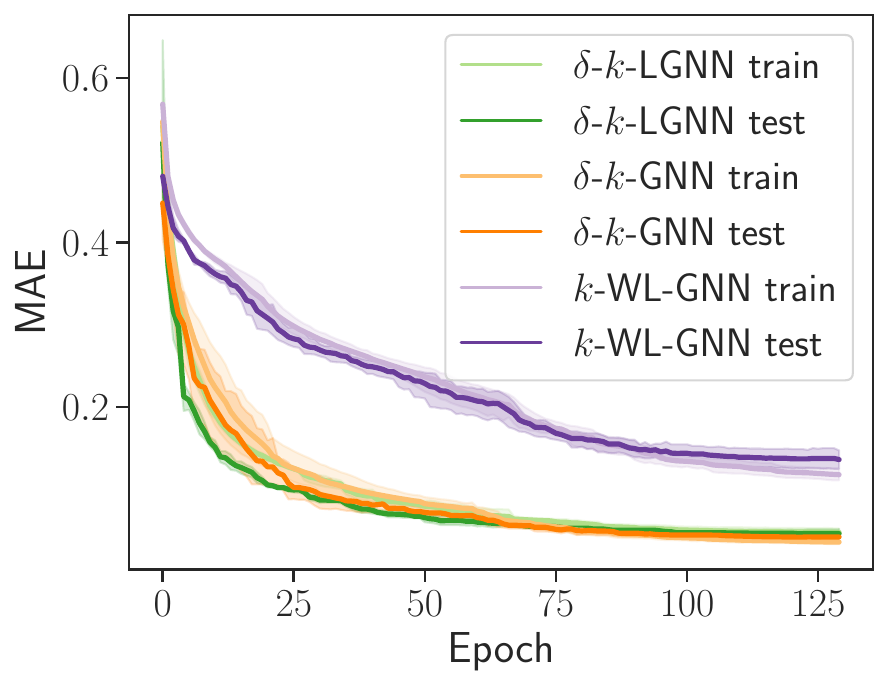}
					\caption{\textsc{Zinc}}
			\label{plot_zinc}	
		\end{subfigure}
		\begin{subfigure}[c]{0.3\textwidth}
			\centering
			\includegraphics[scale=0.28]{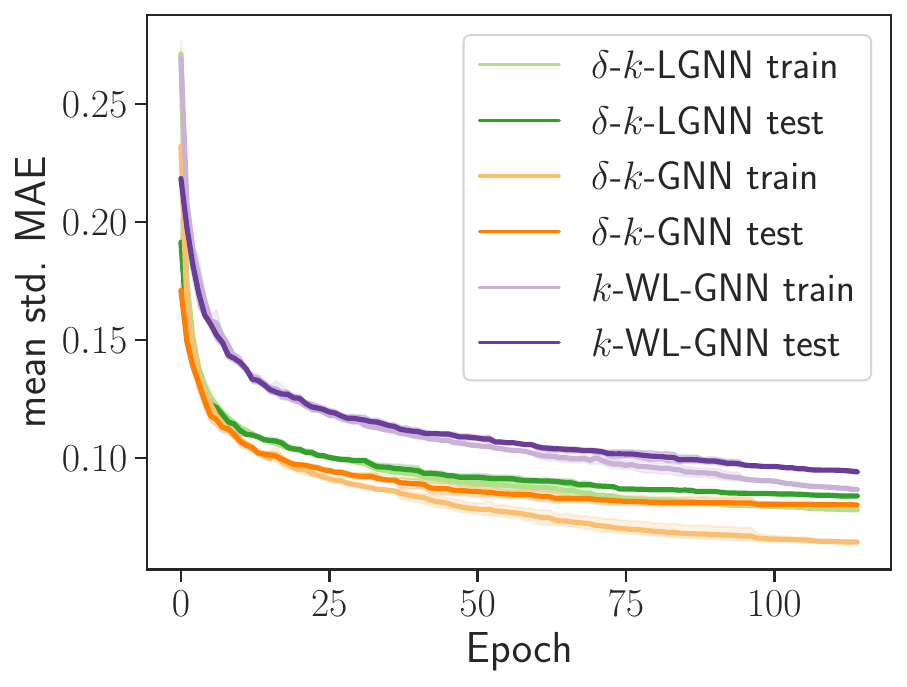}
						\caption{\textsc{Alchemy}}
			\label{plot_alchem}	
		\end{subfigure}
	\end{center}
	\caption{Additional results for neural architectures.}	\label{plot}
\end{figure}

\xhdr{A1} \textit{Kernels} See~\cref{t2}. The local algorithm, for $k=2$  and $3$, severely improves the classification accuracy compared to the \kwl and the \deltakwl. For example, on the \textsc{Enzymes} dataset the $\delta$-$2$-\textsf{LWL} achieves an improvement of almost $20$\%, and the $\delta$-$3$-\textsf{LWL} achieves the best accuracies over all employed kernels, improving over the $3$-\textsf{WL} and the $\delta$-$3$-\textsf{WL} by more than $13$\%. This observation holds over all datasets. Our algorithms also perform better than neural baselines. See~\cref{t2l} in the appendix for additional results on the mid-scale datasets. However, it has to be noted that increasing $k$ does not always result in increased accuracies. For example, on all datasets (excluding \textsc{Enzymes}), the performance of the $\delta$-$2$-\textsf{LWL} is better or on par with the $\delta$-$3$-\textsf{LWL}. Hence, with increasing $k$ the local algorithm is more prone to overfitting.\\
\textit{Neural architectures} See~\cref{neural_short_tt,plot}. On the \textsc{Zinc} and \textsc{Alchemy} datasets, the $\delta$-$2$-\textsf{LGNN} is on par or slightly worse than the $\delta$-$2$-\textsf{GNN}. Hence, this is in contrast to the kernel variant. We assume that this is due to the  $\delta$-$2$-\textsf{GNN} being more flexible than its kernel variant in weighing the importance of global and local neighbors. This is further highlighted by the worse performance of the $2$-\textsf{WL-GNN}, which even performs worse than the ($1$-dimensional) \gineeps on the \textsc{Zinc} dataset. On the \textsc{QM9} dataset, see~\cref{t2nn_short}, the $\delta$-$2$-\textsf{LGNN} performs better than the higher-order methods from~\cite{Mar+2019,Mor+2019} while being on par with the \mpnn architecture. We note here that the \mpnn was specifically tuned to the \textsc{Qm9} dataset, which is not the case for the $\delta$-$2$-\textsf{LGNN} (and the other higher-order architectures).\\
\xhdr{A2} See~\cref{t2}. The $\delta$-$2$-\textsf{LWL}$^+$ improves over the $\delta$-$2$-\textsf{LWL} on all datasets excluding \textsc{Enzymes}. For example, on \textsc{IMDB-Multi}, \textsc{NCI1}, \textsc{NCI109}, and \textsc{Proteins} the algorithm achieves an improvement over of $4\%$, respectively, achieving a new state-of-the-art. The computation times are only increased slightly, see~\cref{t1_app} in the appendix. Similar results can be observed on the mid-scale datasets, see~\cref{t2l,t1l_app} in the appendix.\\
\xhdr{A3} \textit{Kernels} As~\cref{t3_short} (\cref{t3} for all datasets) shows the $\delta$-$2$-\textsf{WL} reaches slightly higher training accuracies over all datasets compared to the $\delta$-$2$-\textsf{LWL}, while the testing accuracies are much lower (excluding \text{PTC\_FM} and \textsc{Proteins}). This indicates that the  $\delta$-$2$-\textsf{WL} overfits on the training set. The higher test accuracies of the local algorithm are likely due to the smaller neighborhood, which promotes that the number of colors grow slower compared to the global algorithm. The \pluskwl inherits the strengths of both algorithms, i.e., achieving the overall best training accuracies while achieving state-of-the-art testing accuracies.\\
\textit{Neural architectures} See~\cref{plot}. In contrast to the kernel variants, the $2$-\textsf{WL} and the $\delta$-$2$-\textsf{WL}, the corresponding neural architectures, the $2$-\textsf{WL-GNN} and the $\delta$-$2$-\textsf{GNN}, seem less prone to overfitting. However, especially on the \textsc{Alchemy} dataset, the $\delta$-$2$-\textsf{LGNN} overfits less. \\
\textbf{A4}
\textit{Kernels} See~\cref{t2lee} (\cref{t1_app,t1l_app} for all datasets). The local algorithm severely speeds up the computation time compared to the \deltakwl and the \kwl for $k=2$ and $3$. For example, on the \textsc{Enzymes} dataset the $\delta$-$2$-\textsf{LWL} is over ten times faster than the $\delta$-$2$-\textsf{WL}.  The improvement of the computation times can be observed across all datasets.  For some datasets, the $\{2,3\}$-\textsf{WL} and the $\delta$-$\{2,3\}$-\textsf{WL} did not finish within the given time limit or went out of memory. For example, on four out of eight datasets, the $\delta$-3-\textsf{WL} is out of time or out of memory. In contrast, for the corresponding local algorithm, this happens only two out of eight times. Hence, the local algorithm is more suitable for practical applications. \\
\textit{Neural architectures} See~\cref{t2len}. The local algorithm severely speeds up the computation time of training and testing. Especially, on the \textsc{Zinc} dataset, which has larger graphs compared to the \textsc{Alchemy} dataset, the $\delta$-$2$-\textsf{LGNN} achieves a computation time that is more  than two times lower compared to the $\delta$-$2$-\textsf{GNN} and the $2$-\textsf{WL-GNN}.
\begin{table}
	\begin{center}
		\begin{subfigure}[c,top]{0.48\textwidth}	\centering
			\caption{\label{t2lee}Speed up ratios of local kernel computations for a subset of the datasets.}
			\resizebox{0.95\textwidth}{!}{\renewcommand{\arraystretch}{1.05}
				\begin{tabular}{@{}c<{\enspace}@{}lccc@{}}
					\toprule
					& \multirow{3}{*}{\vspace*{4pt}\textbf{Method}} &       \multicolumn{3}{c}{\textbf{Dataset}}       \\
					\cmidrule{3-5}               &                                               & {\textsc{Enzymes}} & {\textsc{IMDB-Binary}} & {\textsc{NCI1}} \\ \toprule
					\multirow{2}{*}{\rotatebox{90}{Global\hspace{-1pt}}} &  \textsf{$2$-WL}                                 &        10.4        &   3.6         & 14.3     \\
					& \textsf{$\delta$-$2$-WL}                           &    10.1      &  3.6        &  14.5          \\
					\cmidrule{2-5}      
					& \textsf{$\delta$-$2$-LWL$^+$}                     &          1.2          &           1.2& 1.3            \\
					& \textsf{$\delta$-$2$-LWL}                      &          1.0          &           1.0         &  1.0 \\ \bottomrule
			\end{tabular}}
		\end{subfigure}\hspace{10pt}
		\begin{subfigure}[c,top]{0.48\textwidth}	\centering
			\caption{\label{t2len}Average speed up ratios over all epochs (training and testing).}
			\resizebox{.65\textwidth}{!}{\renewcommand{\arraystretch}{1.05}
				\begin{tabular}{@{}c<{\enspace}@{}lcc@{}}
					\toprule
					& \multirow{3}{*}{\vspace*{4pt}\textbf{Method}} &       \multicolumn{2}{c}{\textbf{Dataset}}       \\
					\cmidrule{3-4}               &                                               & {\textsc{Zinc}} & {\textsc{alchemy}} \\ \toprule
					\multirow{2}{*}{\rotatebox{90}{Dense\hspace{-1pt}}} 
					& \textsf{$2$-WL-GNN}                           &          2.2          &           1.1            \\
					& \textsf{$\delta$-$2$-GNN}                     &          2.5          &           1.7            \\
					\cmidrule{2-4}      
					& \gineeps                                 &          0.2          &           0.4            \\
					& \textsf{$\delta$-$2$-LGNN}                      &          1.0          &           1.0            \\ \bottomrule
			\end{tabular}}
			\vspace{0pt}
		\end{subfigure}
	\end{center}
	\caption{Speed up ratios of local over global algorithms.}\label{add}
\end{table}
\section{Conclusion}
We introduced local variants of the $k$-dimensional Weisfeiler-Leman algorithm. We showed that one variant and its corresponding neural architecture are strictly more powerful than the \kwl while taking the underlying graph's sparsity into account. To demonstrate the practical utility of our findings, we applied them to graph classification and regression. We verified that our local, sparse algorithms lead to vastly reduced computation times compared to their global, dense counterparts while establishing new state-of-the-art results on a wide range of benchmark datasets. \emph{We believe that our local, higher-order kernels and GNN architectures should become a standard approach in the regime of supervised learning with small graphs, e.g., molecular learning.} 

Future work includes a more fine-grained analysis of the proposed algorithm, e.g., moving away from the restrictive graph isomorphism objective and deriving a deeper understanding of the neural architecture's capabilities when optimized with stochastic gradient descent.

\section*{Broader impact}
We view our work mainly as a methodological contribution. It studies the limits of current (supervised) graph embeddings methods, commonly used in chemoinformatics~\cite{Sto+2020}, bioinformatics~\cite{Barabasi2004}, or network science~\cite{Eas+2010}. Currently, methods used in practice, such as GNNs or extended-connectivity fingerprints~\cite{Rogers2010} have severe limitations and might miss crucial patterns in today's complex, interconnected data. We investigate how to scale up graph embeddings that can deal with higher-order interactions of vertices (or atom of molecules, users in social networks, variables in optimization, \dots) to larger graphs or networks.  Hence, our method paves the way for more resource-efficient and expressive graph embeddings.

We envision that our (methodological) contributions enable the design of more expressive and scalable graph embeddings in fields such as quantum chemistry, drug-drug interaction prediction, in-silicio, data-driven drug design/generation, and network analysis for social good. However, progress in graph embeddings might also trigger further advancements in hostile social network analysis, e.g., extracting more fine-grained user interactions for social tracking.

\xhdr{Example impact} We are actively cooperating with chemists on drug design to evaluate further our approach to new databases for small molecules. Here, the development of new databases is quite tedious, and graph embeddings can provide hints to the wet lab researcher where to start their search. However, still, humans need to do much of the intuition-driven, manual wet lab work.  Hence, we do not believe that our methods will result in job losses in the life sciences in the foreseeable future.

\begin{ack}
We thank Matthias Fey for answering our questions with regard to \textsc{PyTorch Geometric} and Xavier Bresson for providing the \textsc{Zinc} dataset.
This work is funded by the Deutsche Forschungsgemeinschaft (DFG, German Research Foundation) under Germany's Excellence Strategy -- EXC-2047/1 -- 390685813
and under DFG Research Grants Program--RA 3242/1-1--411032549. 
\end{ack}

\bibliographystyle{plainnat}

\newpage
\appendix
\section*{Appendix}

\section{Related work (Expanded)}\label{exprel}

In the following, we review related work from graph kernels, GNNs, and theory.

\xhdr{Graph kernels}
Historically, kernel methods---which implicitly or explicitly map graphs to elements of a Hilbert space---have been the dominant approach for supervised learning on graphs. Important early work in this area includes random-walk based kernels~\cite{Gae+2003,Kas+2003,Kri+2017b} and kernels based on shortest paths~\cite{Bor+2005}. More recently, graph kernels' developments have emphasized scalability, focusing on techniques that bypass expensive Gram matrix computations by using explicit feature maps, see, e.g.,~\cite{She+2011}. \citeauthor{Mor+2017}~\cite{Mor+2017} devised a local, set-based variant of the \kwl. However, the approach is (provably) weaker than the tuple-based algorithm, and they do not prove convergence to the original algorithm. \citeauthor{Yan+2015} successfully employed Graphlet~\cite{She+2009}, and Weisfeiler-Leman kernels within frameworks for smoothed~\cite{Yan+2015} and deep graph kernels~\cite{Yan+2015a}. Other recent works focus on assignment-based~\cite{Joh+2015,Kri+2016,Nik+2017}, spectral~\cite{Kon+2016,Ver+2017}, graph decomposition ~\cite{Nik+2018}, randomized binning approaches~\cite{Hei+2019}, and the extension of kernels based on the \wl~\cite{Rie+2019,Tog+2019}. For a theoretical investigation of graph kernels, see~\cite{Kri+2018}, for a thorough survey of graph kernels, see~\cite{Kri+2019}. 

\xhdr{GNNs}
Recently, graph neural networks (GNNs)~\cite{Gil+2017,Sca+2009} emerged as an alternative to graph kernels. Notable instances of this architecture include, e.g.,~\cite{Duv+2015,Fey+2018,Ham+2017,Vel+2018}, and the spectral approaches proposed in, e.g.,~\cite{Bru+2014,Def+2015,Kip+2017,Mon+2017}---all of which descend from early work in~\cite{Kir+1995,Mer+2005,Spe+1997,Sca+2009}. Recent extensions and improvements to the GNN framework include approaches to incorporate different local structures (around subgraphs), e.g.,~\cite{Hai+2019,Fla+2020,Jin+2020,Nie+2016,Xu+2018}, novel techniques for pooling node representations in order perform graph classification, e.g.,~\cite{Can+2018,Gao+2019,Yin+2018,Zha+2018}, incorporating distance information~\cite{You+2019}, and non-euclidian geometry approaches~\cite{Cha+2019}. Moreover, recently empirical studies on neighborhood aggregation functions for continuous vertex features~\cite{Cor+2020}, edge-based GNNs leveraging physical knowledge~\cite{And+2019,Kli+2020}, and sparsification methods~\cite{Ron+2020} emerged. \citeauthor{Loukas20}~\cite{Loukas20} and \citeauthor{Sat+2019} studied the limits of GNNs when applied to combinatorial problems. A survey of recent advancements in GNN techniques can be found, e.g., in~\cite{Cha+2020,Wu+2019,Zho+2018}. \citeauthor{Gar+2020}~\cite{Gar+2020} and~\citeauthor{Ver+2019}~\cite{Ver+2019} studied the generalization abilities of GNNs, and~\cite{Du+2019} related wide GNNs to a variant of the neural tangent kernel~\cite{Aro+2019,Jac+2020}. \citeauthor{Mur+2019}~\cite{Mur+2019a,Mur+2019} and~\citeauthor{Sat+2020}~\cite{Sat+2020} extended the expressivity of GNNs by considering all possible permutations of a graph’s adjacency matrix, or adding random node features, respectivley. The connection between random colorings and universality was investigated in~\cite{Das+2020}.

Recently, connections to Weisfeiler-Leman type algorithms have been shown~\cite{Bar+2020,Che+2019,Gee+2020a,Gee+2020b,Mae+2019,Mar+2019,Mor+2019,Xu+2018b}. Specifically,~\cite{Mor+2019,Xu+2018b} showed that the expressive power of any possible GNN architecture is limited by the \wl in terms of distinguishing non-isomorphic graphs. \citeauthor{Mor+2019}~\cite{Mor+2019} also introduced \emph{$k$-dimensional GNNs} (\kgnn) which rely on a message-passing scheme between subgraphs of cardinality $k$. Similar to~\cite{Mor+2017}, the paper employed a local, set-based (neural) variant of the \kwl, which is (provably) weaker than the variant considered here. Later, this was refined in~\cite{Mar+2019} by introducing \emph{$k$-order invariant graph networks} (\kign), based on \citeauthor{Mar+2019b}~\cite{Mar+2019b}, and references therein, which are equivalent to the folklore variant of the \kwl~\cite{Gro2017} in terms of distinguishing non-isomorphic graphs. However, \kign may not scale since they rely on dense linear algebra routines. \citeauthor{Che+2019}~\cite{Che+2019} connect the theory of universal approximation of permutation-invariant functions and the graph isomorphism viewpoint and introduce a variation of the $2$-\textsf{WL}, which is more powerful than the former. Our comprehensive treatment of higher-order, sparse, neural networks for arbitrary $k$ subsumes all of the algorithms and neural architectures mentioned above.

Finally, there exists a new line of work focusing on extending GNNs to hypergraphs, see, e.g.,~\cite{Bai+2019,Yad+2019,Zha+2019}, and a line of work in the data mining community incorporating global or higher-order information into graph or node embeddings, see, e.g.,~\cite{Cao+2015,Lee+2019,Cha+2018}.

\xhdr{Theory}
The Weisfeiler-Leman algorithm constitutes one of the earliest approaches to isomorphism testing~\cite{Wei+1976,Wei+1968}, having been heavily investigated by the theory community over the last few decades~\cite{Gro+2014}. Moreover, the fundamental nature of the \kwl is evident from a variety of connections to other fields such as logic, optimization, counting complexity, and quantum computing. The power and limitations of $k$-WL can be neatly characterized in terms of logic and descriptive complexity~\cite{Imm+1990}, Sherali-Adams relaxations of the natural integer linear program for the graph isomorphism problem~\cite{Ast+2013,GroheO15,Mal2014}, homomorphism counts~\cite{Del+2018}, and quantum isomorphism games~\cite{Ats+2019}. In their seminal paper~\cite{Imm+1990},~\citeauthor{Cai+1992} showed that for each $k$ there exists a pair of non-isomorphic graphs of size $\cO(k)$ each that cannot be distinguished by the \kwl. \citeauthor{Gro+2014}~\cite{Gro+2014} gives a thorough overview of these results. For $k=1$, the power of the algorithm has been completely characterized~\cite{Arv+2015,Kie+2015}.  Moreover, upper bounds on the running time for $k=1$~\cite{Ber+2013,Kie+2020}, and the number of iterations for the folklore $k=2$~\cite{Kie+2016,Lic+2019} have been shown. For $k=1$ and $2$,~\citeauthor{Arv+2019}~\cite{Arv+2019} studied the abilities of the (folklore) \kwl to detect and count fixed subgraphs, extending the work of~\citeauthor{Fue+2017}~\cite{Fue+2017}. The former was refined in~\cite{Che+2020a}. The algorithm (for logarithmic $k$) plays a prominent role in the recent result of Babai~\cite{Bab+2016} improving the best-known running time for the graph isomorphism problem. Recently,~\citeauthor{Gro+2020a}~\cite{Gro+2020a} introduced the framework of Deep Weisfeiler Leman algorithms, which allow the design of a more powerful graph isomorphism test than Weisfeiler-Leman type algorithms. Finally, the emerging connections between the Weisfeiler-Leman paradigm and graph learning are described in a recent survey of \citeauthor{Gro+2020}~\cite{Gro+2020}.

\section{Preliminaries (Expanded)}\label{prelim}

We briefly describe the Weisfeiler-Leman algorithm and, along the way, introduce our notation. We also state a variant of the algorithm, introduced in~\cite{Mal2014}. As usual, let $[n] = \{ 1, \dotsc, n \} \subset \NN$ for $n \geq 1$, and let $\{\!\!\{ \dots\}\!\!\}$ denote a multiset. 

\xhdr{Graphs} A \new{graph} $G$ is a pair $(V,E)$ with a \emph{finite} set of
\new{vertices} $V$ and a set of \new{edges} $E \subseteq \{ \{u,v\}
\subseteq V \mid u \neq v \}$. We denote the set of vertices and the set
of edges of $G$ by $V(G)$ and $E(G)$, respectively. For ease of
notation, we denote the edge $\{u,v\}$ in $E(G)$ by $(u,v)$ or
$(v,u)$. In the case of \emph{directed graphs} $E \subseteq \{ (u,v)
\in V \times V \mid u \neq v \}$. A \new{labeled graph} $G$ is a triple
$(V,E,l)$ with a label function $l \colon V(G) \cup E(G) \to \Sigma$,
where $\Sigma$ is some finite alphabet. Then $l(v)$ is a
\new{label} of $v$ for $v$ in $V(G) \cup E(G)$. 
The \new{neighborhood} 
of $v$ in $V(G)$ is denoted by $\delta(v) = N(v) = \{ u \in V(G) \mid (v, u) \in E(G) \}$. 
Moreover, its complement $\ndelta(v) = \{ u \in V(G) \mid (v, u) \notin E(G) \}$. 
Let $S \subseteq
V(G)$ then $G[S] = (S,E_S)$ is the \new{subgraph induced} by $S$ with
$E_S = \{ (u,v) \in E(G) \mid u,v \in S \}$. A \new{tree} is a connected graph without
cycles. A \new{rooted tree} is a tree with a designated vertex called \new{root} in which the edges are directed in such a way that they point away from the root. 
Let $p$ be a vertex in a directed tree then we call its out-neighbors \new{children} with parent $p$. 

We say that two graphs $G$ and $H$
are \new{isomorphic} if there exists an edge preserving bijection
$\varphi \colon V(G) \to V(H)$, i.e., $(u,v)$ is in $E(G)$ if and only if
$(\varphi(u),\varphi(v))$ is in $E(H)$. If $G$ and $H$ are isomorphic,
we write $G \simeq H$ and call $\varphi$ an \new{isomorphism} between
$G$ and $H$. Moreover, we call the equivalence classes induced by
$\simeq$ \emph{isomorphism types}, and denote the isomorphism type of $G$ by
$\tau_G$. In the case of labeled graphs, we additionally require that
$l(v) = l(\varphi(v))$ for $v$ in $V(G)$ and $l((u,v)) = l((\varphi(u), \varphi(v)))$ for $(u,v)$ in $E(G)$. 
Let $\vec{v}$ be a \emph{tuple} in $V(G)^k$ for $k > 0$, then $G[\vec{v}]$ is the subgraph induced by the components of $\vec{v}$, where the vertices are labeled with integers from $\{ 1, \dots, k \}$ corresponding to indices of $\vec{v}$. 

\xhdr{Kernels} A \emph{kernel} on a non-empty set $\mathcal{X}$ is a positive semidefinite function 
$k \colon \mathcal{X} \times \mathcal{X} \to \mathbb{R}$.
Equivalently, a function $k$ is a kernel if there is a \emph{feature map} 
$\phi \colon \mathcal{X} \to \mathcal{H}$ to a Hilbert space $\mathcal{H}$ with inner product 
$\langle \cdot, \cdot \rangle$, such that 
$k(x,y) = \langle \phi(x),\phi(y) \rangle$ for all $x$ and $y$ in $\mathcal{X}$.
Let $\mathcal{G}$ be the set of all graphs, then a (positive semidefinite) function $\mathcal{G} \times \mathcal{G} \to \mathbb{R}$ is called a \emph{graph kernel}.

\section{Vertex refinement algorithms (Expanded)}\label{vr}

Let $k$ be a fixed positive integer. As usual, let $V(G)^k$ denote the set of $k$-tuples of vertices of $G$. 

A \new{coloring} of $V(G)^k$ is a mapping $C \colon V(G)^k \to \mathbb{N}$, i.e., we assign a number (color) to every tuple in $V(G)^k$. The \new{initial coloring} $C_{\,0}$ of $V(G)^k$ is specified by the isomorphism types of the tuples, i.e., two tuples $\vec{v}$ and $\vec{w}$ in $V(G)^k$ get a common color iff the mapping $v_i \to w_i$ induces an isomorphism between the labeled subgraphs $G[\vec{v}]$ and $G[\vec{w}]$. A \new{color class} corresponding to a color $c$ is the set of all tuples colored $c$,
i.e., the set $C^{-1}(c)$. 

The \new{neighborhood} of a vertex tuple $\vec{v}$ in $V(G)^k$ is defined as follows. For $j$ in $[k]$, let $\phi_j(\vec{v},w)$ be the $k$-tuple obtained by replacing the 
$j^{\textrm{th}}\!$ component of $\vec{v}$ with the vertex $w$. That is, $\phi_j(\vec{v},w) = (v_1, \dots, v_{j-1}, w, v_{j+1}, \dots, v_k)$. If $\vec{w} = \phi_j(\vec{v},w)$ for some $w$ in $V(G)$, call $\vec{w}$ a $j$-\new{neighbor} of $\vec{v}$. The neighborhood of $\vec{v}$ is thus defined as the set of all tuples $\vec{w}$ such that $\vec{w} = \phi_j(\vec{v},w)$ for some $j$ in $[k]$ and $w$ in $V(G)$. 

The \emph{refinement} of a coloring $C \colon V(G)^k \to \mathbb{N}$, denoted by $\widehat{C}$, is a coloring $\widehat{C} \colon V(G)^k \to \mathbb{N}$ defined as follows. 
For each $j$ in $[k]$, collect the colors of the $j$-neighbors of $\vec{v}$ as a multiset $S_j=\{\!\! \{  C(\phi_j(\vec{v},w)) \mid w \in V(G) \} \!\!\}$.
Then, for a tuple $\vec{v}$, define
\[
\widehat{C}(\vec{v}) = (C(\vec{v}), M(\vec{v})),
\]
where $M(\vec{v})$ is the $k$-tuple $(S_1,\dots,S_k)$. For consistency, the strings $\widehat{C}(\vec{v})$ thus obtained are lexicographically sorted and renamed as integers. Observe that the new color $\widehat{C}(\vec{v})$ of $\vec{v}$ is solely dictated by the color histogram of its neighborhood. In general, a different mapping $M(\cdot)$ could be used, depending on the neighborhood information that we would like to aggregate. We will refer to a mapping $M(\cdot)$ as an \new{aggregation map}. 

\xhdr{$\boldsymbol{k}$-dimensional Weisfeiler-Leman}\label{wl_app} For $k\geq 2$, the \kwl computes a coloring $C_\infty \colon V(G)^k \to \mathbb{N}$ of a given graph $G$, as follows.\footnote{We define the $1$-\textsf{WL} in the next subsection.} To begin with, the initial coloring $C_0$ is computed. Then, starting with $C_0$, successive refinements $C_{i+1} = \widehat{C_i}$ are computed until convergence. That is,
\[
C_{i+1}(\vec{v}) = (C_i(\vec{v}), M_i(\vec{v})),
\]
where 
\begin{equation}\label{app:mi}
M_i(\vec{v}) =   \big( \{\!\! \{  C_i(\phi_1(\vec{v},w)) \mid w \in V(G) \} \!\!\}, \dots, \{\!\! \{  C_i(\phi_k(\vec{v},w)) \mid w \in V(G) \} \!\!\} \big).
\end{equation}
The successive refinement steps are also called \new{rounds} or \new{iterations}. Since the disjoint union of the color classes form a partition of $V(G)^k$, there must exist a finite $\ell \leq |V(G)|^k$ such that $C_{\ell} = \widehat{C_{\ell}}$. In the end, the \kwl outputs $C_\ell$ as the \emph{stable coloring} $C_\infty$. 

The \kwl \new{distinguishes} two graphs $G$ and $H$ if, upon running the \kwl on their disjoint union $G \,\dot\cup\, H$, there exists a color $c$ in $\mathbb{N}$ in the stable coloring such that the corresponding color class $S_c$ satisfies
\begin{equation*}
|V(G)^k \cap S_c| \neq |V(H)^k \cap S_c|,
\end{equation*}
i.e., there exist an unequal number of $c$-colored tuples in $V(G)^k$ and $V(H)^k$. Hence, two graphs distinguished by the \kwl must be non-isomorphic. 

In fact, there exist several variants of the above defined \kwl. These variants result from the application of different aggregation maps $M(\cdot)$. For example, setting $M(\cdot)$ to be 
\begin{equation*}
M^F(\vec{v}) =  \{\!\! \{ \big( \,C(\phi_1(\vec{v},w)) , \dots,   C(\phi_k(\vec{v},w)) \,\big)  \mid w \in V(G) \} \!\!\},
\end{equation*}
yields a well-studied variant of the \kwl (see, e.g., \cite{Cai+1992}), commonly known as ``folklore'' \kwl in machine learning literature. It holds that the $k$-WL using~\cref{app:mi} is as powerful as the folklore $(k\!-\!1)$-WL~\cite{GroheO15}.

\subsection{$\boldsymbol{\delta}$-Weisfeiler-Leman algorithm} 

\begin{figure}[t]
	\begin{center}
		\begin{subfigure}[c]{0.28\textwidth}
			\centering
			\includegraphics[scale=1]{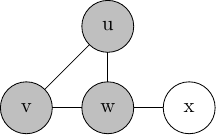}
			\caption{Underlying graph $G$, with tuple $(u,v,w)$}
		\end{subfigure}\hspace{10pt}
		\begin{subfigure}[c]{0.28\textwidth}
			\centering
			\includegraphics[scale=1]{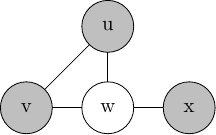}
			\caption{$(u,v,x)$ is a local $3$-neighbor of $(u,v,w)$  }
		\end{subfigure}\hspace{10pt}
		\begin{subfigure}[c]{0.28\textwidth}
			\centering
			\includegraphics[scale=1]{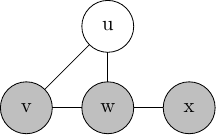}
			\caption{$(x,v,w)$ is a global $1$-neighbor of $(u,v,w)$}
		\end{subfigure}
	\end{center}
	\caption{Illustration of the local and global neighborhood of the $3$-tuple $(u,v,w)$.}\label{kwlill}
\end{figure}

Let $\vec{w} = \phi_j(\vec{v},w)$ be a $j$-{neighbor} of $\vec{v}$. Call $\vec{w}$ a \new{local} $j$-neighbor of $\vec{v}$ if $w$ is adjacent to the replaced vertex $v_j$. Otherwise, call $\vec{w}$ a \new{global} $j$-neighbor of $\vec{v}$. \cref{kwlill} illustrates this definition for a 3-tuple $(u,v,w)$.
For tuples $\vec{v}$ and $\vec{w}$ in $V(G)^k$, the function
\begin{equation*}
\textrm{adj}((\vec{v},\vec{w})) =
\begin{cases} 
\text{L} & \text{if $\vec{w}$ is a local neighbor of $\vec{v}$} \\ 
\text{G} & \text{if $\vec{w}$ is a global neighbor of $\vec{v}$}
\end{cases} 
\end{equation*}
indicates whether $\vec{w}$ is a local or global neighbor of $\vec{v}$. 

The \new{$\delta$-$k$-dimensional Weisfeiler-Leman algorithm}, denoted by \deltakwl, is a variant of the classic \kwl which \emph{differentiates} between the local and the global neighbors during neighborhood aggregation \cite{Mal2014}. Formally, the \deltakwl algorithm refines a coloring $C_i$ (obtained after $i$ rounds) via the aggregation map 
\begin{equation}\label{app:mid}
\begin{split}
M^{\delta,\ndelta}_i(\vec{v}) =   \big( &\{\!\! \{  (C_i({\phi_1(\vec{v},w)}, \textrm{adj}(\vec{v}, \phi_1(\vec{v},w)     )) \mid w \in V(G) \} \!\!\}, \dots, \\ &\{\!\! \{  (C_i({\phi_k(\vec{v},w)}, \textrm{adj}(\vec{v},  \phi_k(\vec{v},w)    )) \hspace{-0.5pt}\mid w \in V(G) \}  \!\!\} \big),
\end{split}	
\end{equation}
instead of the \kwl aggregation specified by~\cref{app:mi}. We define the $1$-\textsf{WL} to be the $\delta$-1-\textsf{WL}, which is commonly known as color refinement or naive vertex classification.

\xhdr{Comparing $\boldsymbol{k}$-\textsf{WL} variants} Given that there exist several variants ofthe \kwl, corresponding to different aggregation maps $M(\cdot)$, it is natural to ask whether they are equivalent in power, vis-a-vis distinguishing non-isomorphic graphs. Let $A_1$ and $A_2$ denote two vertex refinement algorithms, we write
$ A_1 \sqsubseteq A_2$ if $A_1$ distinguishes between all non-isomorphic pairs $A_2$ does, and $A_1 \equiv A_2$ if both directions hold. The corresponding strict relation is denoted by $\sqsubset$. 

The following result relates the power of the \kwl and \deltakwl. Since for a graph $G=(V,E)$,  $M^{\delta,\ndelta}_i(\vec{v}) = M^{\delta,\ndelta}_i(\vec{w})$ implies $M_i(\vec{v}) = M_i(\vec{w})$  for all $\vec{v}$ and $\vec{w}$ in $V(G)^k$ and $i\geq 0$, it immediately follows that $\deltakwlm \sqsubseteq \kwlm$. 
For $k=1$, these two algorithms are equivalent by definition. For $k \geq 2$, this relation can be shown to be strict, see the next section.

\begin{proposition}[restated,~\cref{rel1} in the main text]\label{app:rel1} For all graphs and $k \geq 2 $, the following holds: 	\begin{equation*}
	\deltakwlm \sqsubset \kwlm.
	\end{equation*}
\end{proposition}

\subsubsection{Proof of \cref{rel1}}\label{proofrel1}

It suffices to show an infinite family of graphs $(G_k,H_k)$, $k \in \mathbb{N}$, such that (a) \kwl does not distinguish $G_k$ and $H_k$, although (b) 
\deltakwl distinguishes $G_k$ and $H_k$. 

We proceed to the construction of this family. The graph family is based on the classic construction of \cite{Cai+1992}, commonly referred to as Cai-Furer-Immermman (CFI) graphs. 

\xhdr{Construction.} Let $K$ denote the complete graph on $k+1$ vertices (there are no loops in $K$). The vertices of $K$ are numbered from $0$ to $k$. Let $E(v)$ denote the set of edges incident to $v$ in $K$: clearly, $|E(v)| = k$ for all $v \in V(K)$. Define the graph $G$ as follows:
\begin{enumerate}
	\item For the vertex set $V(G)$, we add   
	\begin{enumerate}
		\item[(a)] $(v,S)$ for each $v \in V(K)$ and for each \emph{even} subset $S$ of $E(v)$, 
		\item[(b)] two vertices $e^1,e^0$ for each edge $e \in E(K)$.  
	\end{enumerate} 
	\item For the edge set $E(G)$, we add 
	\begin{enumerate}
		\item[(a)] an edge $\{e^0,e^1\}$ for each $e \in E(K)$, 
		\item[(b)] an edge between $(v,S)$ and $e^1$ if $v \in e$ and $e \in S$,  
		\item[(c)] an edge between $(v,S)$ and $e^0$ if $v \in e$ and $e \not S$,  
	\end{enumerate} 
\end{enumerate} 
Define a companion graph $H$, in a similar manner to $G$, with the following exception: in Step 1(a), for the vertex $0 \in V(K)$, we choose all \emph{odd} subsets of $E(0)$. Counting vertices, we find that $|V(G)| = |V(H)| = (k+1)\cdot 2^{k-1} + {k \choose 2} \cdot 2$. This finishes the construction of graphs $G$ and $H$. We set $G_k := G$ and $H_k := H$. 

A set $S$ of vertices is said to form a \emph{distance-two-clique} if the distance between any two vertices in $S$ is exactly two. 
\begin{lemma}
	The following holds for graphs $G$ and $H$ defined above. 
	\begin{itemize}
		\item There exists a distance-two-clique of size $(k+1)$ inside $G$.
		\item There does not exist a distance-two-clique of size $(k+1)$ inside $H$.
	\end{itemize}
	Hence, $G$ and $H$ are non-isomorphic. 
\end{lemma}
\begin{proof}
	In the graph $G$, consider the vertex subset $S = \{(0,\emptyset), (1,\emptyset), \dots, (k,\emptyset)\}$ of size $(k+1)$. That is, from each ``cloud'' of vertices of the form $(v,S)$ for a fixed $v$, we pick the vertex corresponding to the trivial even subset, the empty set denoted by $\emptyset$. Observe that any two vertices in $S$ are at distance two from each other. This holds because for any $i,j \in V(K)$,
	$(i,\emptyset)$ is adjacent to $\{i,j\}^0$ which is adjacent to $(j,\emptyset)$ (e.g. see Figure 1). Hence, the vertices in $S$ form a distance-two-clique of size $k+1$. 
	
	On the other hand, for the graph $H$, suppose there exists a distance-two-clique, say $(0,S_0),\dots, (k,S_k)$ in $H$,
	where each $S_i \subseteq E(i)$. If we compute the parity-sum of the parities of $|S_0|,\dots,|S_k|$, we end up with $1$ since there is exactly one odd subset in this collection, viz. $S_0$. On the other hand, we can also compute this parity-sum in an edge-by-edge manner: for each edge $(i,j)\in E(K)$, since $(i,S_i)$ and $(j,S_j)$ are at distance two, either both $S_i$ and $S_j$ contain the edge $\{i,j\}$ or neither of them contains $\{i,j\}$: hence, the parity-sum contribution of $S_i$ and $S_j$ to the term corresponding to $e$ is zero. Since the contribution of each edge to the total parity-sum is $0$, the total parity-sum must be zero. This is a contradiction, and hence, there does not exist a distance-two-clique in $H$. 
\end{proof}  

Next, we show that the local algorithm \localkwl can distinguish $G$ and $H$. Since $\deltakwlm \sqsubseteq$ \localkwl, the above lemma implies the strictness condition $\deltakwlm \sqsubset \kwlm$. 

\begin{lemma}
	\localkwl distinguishes $G$ and $H$. 
\end{lemma} 
\begin{proof}
	The proof idea is to show that \localkwl algorithm is powerful enough to detect distance-two-cliques of size $(k+1)$, which ensures the distinguishability of $G$ and $H$. Indeed, consider the $k$-tuple $P$ = $( (1,\emptyset), (2,\emptyset) , \dots, (k,\emptyset))$ in $V(G)^k$. We claim that there is no tuple $Q$ in $V(H)^k$ such that the unrolling of $P$ is isomorphic to the unrolling of $Q$. Indeed, for the sake of contradiction, assume that there does exist $Q$ in $V(H)^k$ such that the unrolling of $Q$ is isomorphic to the unrolling of $P$. Comparing isomorphism types, we know that the tuple $Q$ must be of the form $((1,S_1), \dots, (k,S_k))$. 
	
	Consider the depth-two unrolling of $P$: from the root vertex $P$, we can go down via two local-edges labeled $1$, to hit the tuple $P_2=((2,\emptyset),(2,\emptyset), \dots, (k,\emptyset))$. If we consider the depth-two unrolling of $Q$, the isomorphism type of $P_2$ implies that the vertices $(1,S_1)$ and $(2,S_2)$ must be at distance-two in the graph $H$. Repeating this argument, we obtain that $(1,S_1), \dots,(k,S_k)$ form a distance-two-clique in $H$ of size $k$. Our goal is to produce a distance-two-clique in $H$ of size $k$, for the sake of contradiction. 
	
	For that, consider the depth-four unrolling of $P$: from the root vertex $P$, we can go down via two local-edges labeled $1$ to hit the tuple $R = ((0,\emptyset), (2,\emptyset), \dots (k,\emptyset)$. For each $2 \leq j \leq k$, we can further go down from $R$ via two local edges labeled $j$ to reach a tuple whose $1^{\text{st}}$ and $j^{\text{th}}$ entry is $(0,\emptyset)$. Similarly, for the unrolling of $Q$, there exists a subset $S_0 \subseteq E(0)$ and a corresponding tuple $R'=((0,S_0),(2,S_2),\dots, (k,S_k))$, such that for each $2 \leq j \leq k$, we can further go down from $R'$ via two local edges labeled $j$ to reach a tuple whose $1^{\text{st}}$ and $j^{\text{th}}$ entry is $(0,S_0)$. Comparing the isomorphism types of all these tuples, we deduce that $(0,S_0)$ must be at distance two from each of $(i,S_i)$ for $i \in [k]$. This implies that the vertex set $\{ (0,S_0),(1,S_1),\dots,(k,S_k)\}$ is a distance-two-clique of size $k+1$ in $H$, which is impossible. Hence, there does not exist any $k$-tuple $Q$ in $V(H)^k$ such that the unrolling of $P$ and the unrolling of $Q$ are isomorphic. Hence, the \localkwl distinguishes $G$ and $H$. 
\end{proof}

Finally, we note that CFI graphs are standard tools from graph isomorphism theory, and are often used to analyze the power and limitations of WL-type algorithms. It follows from results of \cite{Cai+1992} that for every $k \geq 0$, \kwl fails to distinguish the graphs $G_k$ and $H_k$ of our constructed family. This finishes the proof of the proposition. 

\section{Local $\boldsymbol{\delta}$-$\boldsymbol{k}$-dimensional Weisfeiler-Leman algorithm (Expanded)}\label{lwl}

In this section, we define the new \new{local $\delta$-$k$-dimensional Weisfeiler-Leman algorithm} (\localkwl). This variant of \deltakwl considers only local neighbors during the neighborhood aggregation process, and discards any information about the global neighbors. Formally, the \localkwl algorithm refines a coloring $C_i$ (obtained after $i$ rounds) via the aggregation map, 
\begin{equation}\label{app:eqnmidd}
\begin{split}
M^{\delta}_i(\vec{v}) =   \big( \{\!\! \{ C_{i}(\phi_1(\vec{v},w)) \mid w \in N(v_1) \} \!\!\}, \dots, \{\!\! \{  C_{i}(\phi_k(\vec{v},w)) \mid w \in N(v_k) \}  \!\!\} \big),
\end{split}
\end{equation}		
instead of~\cref{app:mid}. That is, the algorithm only considers the local $j$-neighbors of the vertex $\vec{v}$ in each iteration. 
Therefore, the indicator function $\mathrm{adj}$ used in~\cref{app:mid} is trivially equal to $L$ here, and is hence omitted. 
The coloring function for the \localkwl is defined by
\begin{equation*}\label{app:ck}
C^{k,\delta}_{i+1}(\vec{v}) = (C^{k,\delta}_{i}(\vec{v}), M^{\delta}_i(\vec{v})).
\end{equation*} 
We also define \pluskwl, a minor variation of \localkwl. Later, we will show that \pluskwl is equivalent in power to $\delta$-$k$-WL (\cref{app:loco}). 
Formally, the \pluskwl algorithm refines a coloring $C_i$ (obtained after $i$ rounds) via the aggregation function, 
\begin{equation}\label{middp}
\begin{split}
M^{\delta,+}(\vec{v}) =   \big( &\{\!\! \{ (C_i(\phi_1(\vec{v},w)), \#_{i}^1(\vec{v},\phi_1(\vec{v},w))) \mid w \in N(v_1) \} \!\!\}, \dots, \\ &\{\!\! \{  (C_i(\phi_k(\vec{v},w)), \#_{i}^k(\vec{v},\phi_k(\vec{v},w))) \mid w \in N(v_k) \}  \!\!\} \big),
\end{split}
\end{equation}
instead of \localkwl aggregation defined in~\cref{app:eqnmidd}. 
Here, the function
\begin{equation*}
\#_{i}^j(\vec{v}, \vec{x}) = \big|\{ \vec{w} \colon \, \vec{w} \sim_j \vec{v}, \, C_{i}(\vec{w}) = C_{i}(\vec{x})   \} \big|,
\end{equation*}
where $\vec{w} \sim_j \vec{v}$ denotes that $\vec{w}$ is $j$-neighbor of $\vec{v}$, for $j$ in $[k]$. Essentially, $\#_{i}^j(\vec{v}, \vec{x})$ counts the number of $j$-neighbors (local or global) of $\vec{v}$ which have the same color as $\vec{x}$ under the coloring $C_i$ (i.e., after $i$ rounds). For a fixed $\vec{v}$,
the function $\#_{i}^j(\vec{v},\cdot)$ is uniform over the set $S \cap N_j$, where $S$ is a color class obtained after $i$ iterations of the \pluskwl and $N_j$ denotes the set of $j$-neighbors of $\vec{v}$. Note that after the stable partition has been reached $\#_{i}^j(\vec{v})$ will not change anymore. Observe that each iteration of the \pluskwl has the same asymptotic running time as an iteration of the \localkwl.

The following theorem shows that the local variant \pluskwl is at least as powerful as the \deltakwl when restricted to the class of connected graphs. In other words, given two \emph{connected} graphs $G$ and $H$, if these graphs are distinguished by \deltakwl, then they must also be distinguished by the \pluskwl. On the other hand, it is important to note that, in general, the \pluskwl might need a larger number of iterations to distinguish two graphs, as compared to \deltakwl. However, this leads to advantages in a machine learning setting, see~\cref{exp}.
\begin{theorem}[restated,~\cref{loco} in the main text]\label{app:loco}
	For the class of connected graphs, the following holds for all $k \geq 1$:
	\begin{equation*}
	\pluskwlm \equiv \deltakwlm.
	\end{equation*}
\end{theorem}	
Along with~\cref{rel1}, we obtain the following corollary relating the power of \kwl and \pluskwl. 
\begin{corollary}[restated,~\cref{cloco} in the main text] For the class of connected graphs, the following holds for all $k \geq 2$:	\begin{equation*}
	\pluskwlm \sqsubset \kwlm.
	\end{equation*}
\end{corollary}
In fact, the proof of \cref{rel1} shows that the infinite family of graphs $G_k, H_k$ witnessing the strictness condition can even be distinguished by \localkwl,
for each corresponding $k \geq 2$. We note here that the restriction to connected graphs can easily be circumvented by adding a specially marked vertex, which is connected to every other vertex in the graph.

\subsection{Kernels based on vertex refinement algorithms}

The idea for a kernel based on the \localkwl (and the other vertex refinements algorithms) is to compute it for $h \geq 0$ iterations resulting in a coloring function $C^{k,\delta} \colon V(G) \to \Sigma_i$ for each iteration $i$. Now, after each iteration, we compute a \new{feature vector} $\phi_i(G)$ in $\bbR^{|\Sigma_i|}$ for each graph $G$. Each component $\phi_i(G)_{c}$ counts the number of occurrences of $k$-tuples labeled by $c$ in $\Sigma_i$. The overall feature vector $\phi_{\text{LWL}}(G)$ is defined as the concatenation of the feature vectors of all $h$ iterations, i.e., $\phi_{\text{LWL}}(G) = \big[\phi_0(G), \dots, \phi_h(G) \big]$. The corresponding kernel for $h$ iterations then is computed as  $k_{\text{LWL}}(G,H) = \langle \phi_{\text{LWL}}(G), \phi_{\text{LWL}}(H) \rangle$, where $\langle \cdot, \cdot \rangle$ denotes the standard inner product.

\subsection{Local converges to global: Proof of~\cref{loco}}\label{locodes}

The main technique behind the proof is to encode the colors assigned by the \kwl (or its variants) as rooted directed trees, called \emph{unrolling trees}. 
The exact construction of the unrolling tree depends on the aggregation map $M(\cdot)$ used by the \kwl variant under consideration. 
We illustrate this construction for the \kwl. For other variants such as the \deltakwl, \localkwl, and \pluskwl, we will specify analogous constructions. 

\xhdr{Unrollings (``Rolling in the deep'')} Given a graph $G$, a tuple $\textbf{v}$ in $V(G)^k$, and an integer $\ell \geq 0$, 
the \emph{unrolling} $\UNR[G,\vec{v},\ell]$ is a rooted, directed tree with vertex and edge labels, defined recursively as follows. 
\begin{itemize}
	\item[-] For $\ell = 0$, $\UNR[G,\textbf{v},0]$ is defined to be a single vertex, labeled with the isomorphism type $\tau(\vec{v})$.  
	This lone vertex is also the root vertex.
	
	\item[-] For $\ell > 0$, $\UNR[G,\textbf{v},\ell]$ is defined as follows. 
	First, introduce a root vertex $r$, labeled with the isomorphism type $\tau(\vec{v})$.
	Next, for each $j \in [k]$ and for each $j$-neighbor $\vec{w}$ of $\vec{v}$, 
	append the rooted subtree $\UNR[G,\vec{w},\ell-1]$ below the root $r$. 
	Moreover, the directed edge $e$ from $r$ to the root of $\UNR[G,\vec{w},\ell-1]$ is labeled $j$ iff $\vec{w}$ is a $j$-neighbor of $\vec{v}$. 
	
\end{itemize}

\begin{figure}
	
	\centering
	\begin{tikzpicture}[scale=0.8]
	\tikzset{
		vertex/.style = {circle, fill = purple,opacity=0.3, minimum size = 50pt}, empty/.style = {},
		mini/.style = {circle,fill, scale = 0.5},
		minie/.style = {-, thick},
		pre/.style = {<-, shorten <= 1pt, >=stealth', thick}
	}
	\node[vertex] (root) at (0,0) [label = above : $\textcolor{black}{\ensuremath{(u,v,w)}}$] {};
	\begin{scope}[yshift = -0.25cm, scale = 0.7]
	\node[mini] (v1) at (-0.5,0) [label = left: \small 1] {};
	\node[mini] (v2) at (+0.5,0) [label = right: \small 2] {} edge[minie] (v1);
	\node[mini] (v3) at (0,0.85) [label = above: \small 3] {} edge[minie] (v1) edge[minie] (v2);
	\end{scope}
	
	\node[vertex] (w2) at (1.5,-2.5) [label = below : $\textcolor{black}{\ensuremath{(u,u,w)}}$] {} edge[pre] node[auto,swap] {2} (root);
	\begin{scope}[xshift = 1.5cm, yshift = -2.50cm, scale = 0.7]
	\node[mini] (v1) at (-0.5,0) [label = above: \small 1, label = below: \small 2] {};
	\node[mini] (v2) at (+0.5,0) [label = right: \small 3] {} edge[minie] (v1);
	\end{scope}
	
	\node[vertex] (w3) at (-1.5,-2.5) [label = below : $\textcolor{black}{\ensuremath{(u,v,w)}}$] {} edge[pre] node[auto] {2} (root);
	\begin{scope}[xshift = -1.5cm, yshift = -2.75cm, scale = 0.7]
	\node[mini] (v1) at (-0.5,0) [label = left: \small 1] {};
	\node[mini] (v2) at (+0.5,0) [label = right: \small 2] {} edge[minie] (v1);
	\node[mini] (v3) at (0,0.85) [label = above: \small 3] {} edge[minie] (v1) edge[minie] (v2);
	\end{scope}
	
	\node[vertex] (w1) at (4.5,-2.5)[label = below : $\textcolor{black}{\ensuremath{(u,w,w)}}$] {} edge[pre] node[auto,swap] {2} (root);
	\begin{scope}[xshift = 4.5cm, yshift = -2.5cm, scale = 0.7]
	\node[mini] (v1) at (-0.5,0) [label = left: \small 1] {};
	\node[mini] (v2) at (+0.5,0) [label = above: \small 2, label = below: \small 3] {} edge[minie] (v1);
	\end{scope}
	
	\node[vertex] (w4) at (-4.5,-2.5) [label = below : $\textcolor{black}{\ensuremath{(u,x,w)}}$] {} edge[pre] node[auto] {2} (root);
	\begin{scope}[xshift = -4.5cm, yshift = -2.75cm, scale = 0.7]
	\node[mini] (v1) at (-0.5,0) [label = left: \small 1] {};
	\node[mini] (v2) at (+0.5,0) [label = right: \small 2] {};
	\node[mini] (v3) at (0,0.85) [label = above: \small 3] {} edge[minie] (v1) edge[minie] (v2);
	\end{scope}
	
	\node[empty] (left) at (-7.5, -1.5) [label = below: $\textcolor{black}{\ensuremath{(*,v,w)}}$] {} edge[pre,dotted] node[auto] {$1$-nbrs} (root); 
	\node[empty] (right) at (7.5, -1.5) [label = below: $\textcolor{black}{\ensuremath{(u,v,*)}}$] {} edge[pre,dotted] node[auto,swap] { $3$-nbrs} (root);
	
	\end{tikzpicture}
	\caption{Unrolling at the tuple $(u,v,w)$ of depth one.}	\label{rolling_app}
\end{figure}
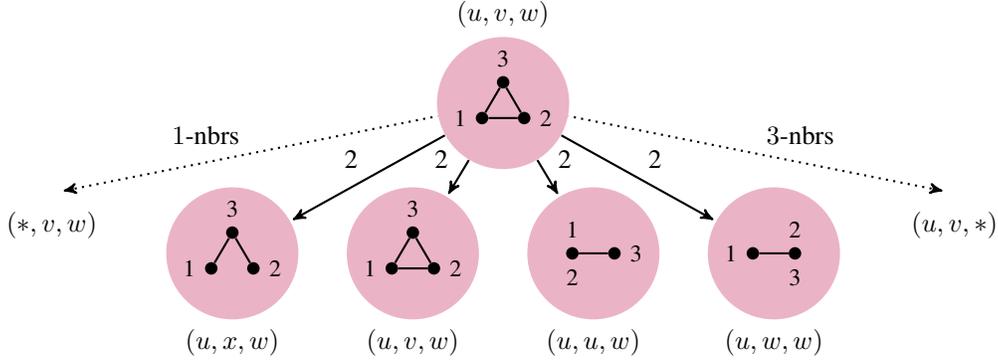

We refer to $\UNR[G,\vec{v},\ell]$ as the unrolling of the graph $G$ \new{at $\vec{v}$ of depth $\ell$}. \cref{rolling_app} partially illustrates the recursive construction of unrolling trees: it describes the unrolling tree for the graph in \cref{kwlill} at the tuple $(u,v,w)$, of depth $1$. Each node $w$ in the unrolling tree is associated with some $k$-tuple $\vec{w}$, indicated alongside the node in the figure. We call $\vec{w}$ the tuple corresponding to the node $w$.

Analogously, we can define unrolling trees $\deltaunr$, $\localunr$, and $\plusunr$ for the \kwl-variants \deltakwl, \localkwl, and \pluskwl respectively. The minor differences lie in the recursive step above, since the unrolling construction needs to faithfully represent the aggregation process. 

\begin{itemize}
	\item[-] For $\deltaunr$, we additionally label the directed edge $e$ with $(j,L)$ or $(j,G)$ instead of just $j$, depending on whether the neighborhood is local or global. 
	\item[-] For $\localunr$, we consider only the subtrees $\localunr[G,\vec{w},\ell-1]$ for local $j$-neighbors $\vec{w}$. 
	\item[-] For $\plusunr$, we again consider only the subtrees $\plusunr[G,\vec{w},\ell-1]$ for local $j$-neighbors $\vec{w}$. However,
	the directed edge $e$ to this subtree is also labeled with the $\#$ counter value $\#_{\ell-1}^j(\vec{v},\vec{w})$.
\end{itemize}

\xhdr{Encoding colors as trees} The following Lemma shows that the computation of the \kwl can be faithfully encoded by the unrolling trees. 
Formally, let $\vec{s}$ and $\vec{t}$ be two $k$-vertex-tuples in $V(G)^k$.

\begin{lemma}\label{encwl}
	The colors of $\vec{s}$ and $\vec{t}$ after $\ell$ rounds of \kwl are identical 
	if and only if the unrolling tree $\UNR[G,\vec{s},\ell]$ is isomorphic to the unrolling tree $\UNR[G,\vec{t},\ell]$.  
\end{lemma}

\begin{proof}
	By induction on $\ell$. For the base case $\ell = 0$, observe that the initial colors of $\vec{s}$ and $\vec{t}$ are equal to the 
	respective isomorphism types $\tau(\vec{s})$ and $\tau(\vec{t})$. On the other hand, the vertex labels for the single-vertex graphs $\UNR[G,\vec{s},0]$ and $\UNR[G,\vec{t},0]$ are also the respective isomorphism types $\tau(\vec{s})$ and $\tau(\vec{t})$. Hence, the statement holds for $\ell = 0$. 
	
	For the inductive case, we proceed with the forward direction. Suppose that \kwl assigns the same color to $\vec{s}$ and $\vec{t}$ after $\ell$ rounds. For each $j$ in $[k]$, the $j$-neighbors of $\vec{s}$ form a partition $\vec{C}_1,\dots,\vec{C}_p$ corresponding to their colors after $\ell-1$ rounds of \kwl. Similarly, the $j$-neighbors of $\vec{t}$ form a partition $\vec{D}_1,\dots,\vec{D}_p$ corresponding to their colors after $\ell-1$ rounds of \kwl, where for $i$ in $[p]$, $\vec{C}_i$ and $\vec{D}_i$ have the same size and correspond to the same color. By inductive hypothesis, the corresponding depth $\ell -1$ unrollings $\UNR[G,\vec{c},\ell-1]$ and $\UNR[G,\vec{d},\ell-1]$ are isomorphic, for every $\vec{c}$ in $\vec{C}_i$ and $\vec{d}$ in $\vec{D}_i$. Since we have a bijective correspondence between the depth $\ell-1$ unrollings of the $j$-neighbors of $\vec{s}$ and $\vec{t}$, respectively, there exists an isomorphism between $\UNR[G,\vec{s},\ell]$ and $\UNR[G,\vec{t},\ell]$. Moreover, this isomorphism preserves vertex labels (corresponding to isomorphism types) and edges labels (corresponding to $j$-neighbors). 
	
	For the backward direction, suppose that $\UNR[G,\vec{s},\ell]$ is isomorphic to $\UNR[G,\vec{t},\ell]$. Then, we have a bijective correspondence between the depth $\ell-1$ unrollings of the $j$-neighbors of $\vec{s}$ and of $\vec{t}$, respectively. For each $j$ in $[k]$, the $j$-neighbors of $\vec{s}$ form a partition $\vec{C}_1,\dots,\vec{C}_p$ corresponding to their unrolling trees after $\ell-1$ rounds of $k$-WL. Similarly, the $j$-neighbors of $\vec{t}$ form a partition $\vec{D}_1,\dots,\vec{D}_p$ corresponding to their unrolling trees after $\ell-1$ rounds of $k$-WL, where for $i$ in $[p]$, $C_i$, and $D_i$ have the same size and correspond to the same isomorphism type of the unrolling tree. By induction hypothesis, the $j$-neighborhoods of $\vec{s}$ and $\vec{t}$ have an identical color profile after $\ell-1$ rounds. Finally, since the depth $\ell -1$ trees $\UNR[G,\vec{s},\ell-1]$ and $\UNR[G,\vec{t},\ell-1]$ are trivially isomorphic, the tuples $\vec{s}$ and $\vec{t}$ have the same color after $\ell-1$ rounds. Therefore, \kwl must assign the same color to $\vec{s}$ and $\vec{t}$ after $\ell$ rounds.  
\end{proof}

Using identical arguments, we can state the analogue of~\cref{encwl} for the algorithms \deltakwl, \localkwl, \pluskwl, and their corresponding unrolling constructions $\deltaunr$, $\localunr$ and $\plusunr$. The proof is identical and is hence omitted.

\begin{lemma}\label{enc}
	The following statements hold. 
	\begin{enumerate}
		\item The colors of $\vec{s}$ and $\vec{t}$ after $\ell$ rounds of \deltakwl are identical 
		if and only if the unrolling tree $\deltaunr[G,\vec{s},\ell]$ is isomorphic to the unrolling tree $\deltaunr[G,\vec{t},\ell]$.   
		\item The colors of $\vec{s}$ and $\vec{t}$ after $\ell$ rounds of \localkwl are identical 
		if and only if the unrolling tree $\localunr[G,\vec{s},\ell]$ is isomorphic to the unrolling tree $\localunr[G,\vec{t},\ell]$.   
		\item The colors of $\vec{s}$ and $\vec{t}$ after $\ell$ rounds of \pluskwl are identical 
		if and only if the unrolling tree $\plusunr[G,\vec{s},\ell]$ is isomorphic to the unrolling tree $\plusunr[G,\vec{t},\ell]$.   
	\end{enumerate}
\end{lemma}

\xhdr{Equivalence} The following Lemma establishes that the local algorithm \pluskwl is at least as powerful as the global \deltakwl, for connected graphs, i.e., $\pluskwlm \sqsubseteq \deltakwlm$. 

\begin{lemma}\label{pluseqtodelta}
	Let $G$ be a connected graph, and let $\vec{s}$ and $\vec{t}$ in $V(G)^k$. If the stable colorings of $\vec{s}$ and $\vec{t}$ under the \pluskwl are identical, 
	then the stable colorings of $\vec{s}$ and $\vec{t}$ under \deltakwl are also identical. 
\end{lemma}

\begin{figure}
	\centering

	\begin{tikzpicture}[scale=0.75]
	\tikzset{
		empty/.style = {},
		vertex/.style = {circle, draw, fill=white, scale = 0.8},
		node/.style = {circle, draw = red!20, fill = red!20},
		pre/.style = {<-, shorten <= 1pt, >=stealth', semithick},
		predot/.style = {<-, shorten <= 1pt, >=stealth', dotted, semithick},
		post/.style = {->, shorten >= 1pt, >=stealth', semithick},
		undir/.style = {-, shorten <= 1pt, >=stealth', dotted, thick},
		prepost/.style = {<->, shorten <= 1pt, >=stealth', dotted, thick}
	}
	\fill[black!10!white] (0,0) -- (-3,-8) -- (3,-8) -- (0,0);
	\node[vertex] (s) [label = left : $\vec{s}\,\,$] {$s$}; 
	
	\fill[black!20!white] (0.3,-3) -- (0.3-1.9,-8) -- (0.3+1.9,-8) -- (0.3,-3);
	\node[vertex] (w) at (0.3,-3) [label = left:$\vec{w}\,\,$] {$w$}; 
	
	\fill[black!40!white] (0.5,-4.5) -- (0.5-1,-8) -- (0.5+1,-8) -- (0.5,-4.5);
	\node[vertex] (x) at (0.5,-4.5) [label = left:$\vec{x}\,$] {$x$} edge [pre,thick] node[midway,right] {$(j,\#)$} (w); 
	
	\draw [->,decorate,decoration={snake,amplitude=1.2mm,segment length=3mm,post length=1mm}]
	(s) -- (w) node [below,align=center,midway] (lab) {$j$ \quad};
	
	\fill[black!10!white] (9+0,0) -- (9-3,-8) -- (9+3,-8) -- (9+0,0);
	\node[vertex] (t) at (9,0) [label = right:$\vec{t}$] {$t$}; 
	
	\fill[black!20!white] (+0.3+9,-3) -- (0.3-1.9+9,-8) -- (0.3+1.9+9,-8) -- (0.3+9,-3);
	\node[vertex] (z) at (+0.3+9,-3) [label = right: $\vec{z}\,$] {$z$}; 
	
	\fill[black!40!white] (+9,-4.5) -- (-1+9,-8) -- (+1+9,-8) -- (+9,-4.5);
	\node[vertex] (y) at (+9,-4.5) [label = right: $\vec{y}$] {$y$} edge [pre,thick] node[midway,left] {$(j,\#)$} (z); 
	
	\draw [->,decorate,decoration={snake,amplitude=1.2mm,segment length=3mm,post length=1mm}]
	(t) -- (z) node [below,align=center,midway] (lab2) {$j$ \quad};

	\draw [post,dashed] (s) to node[midway, label={[shift={(0.2,0)}] $\theta$}] {} (t) ;
	\draw [post,dashed] (w) to node[midway, label= $\theta$] {} (z) ;
	\draw [post,dashed] (x) to node[midway, label={[shift={(0.1,0)}] $\theta$}] {} (y) ;
	
	\draw [prepost] (1.5,-4.5) to node {\quad \quad \quad $> r^*$ } (1.5,-8);
	\end{tikzpicture}
	\caption{Unrollings $L_1 = \plusunr[G,\vec{s},{q}]$ and $L_2 = \plusunr[G,\vec{t},{q}]$ of sufficiently large depth.}\label{rollrollmap}
\end{figure}
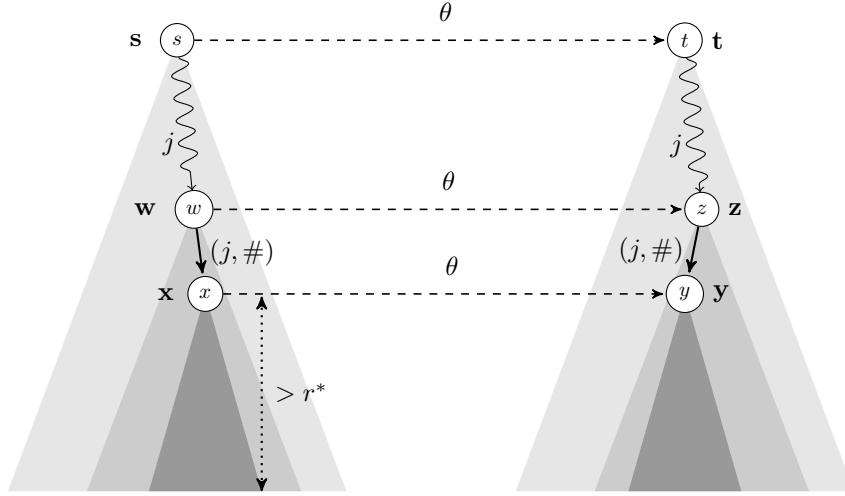

\begin{proof}  
	Let $r^*$ denote the number of rounds needed to attain the stable coloring under \pluskwl. Consider unrollings $L_1 = \plusunr[G,\vec{s},{q}]$ and $L_2 = \plusunr[G,\vec{t},{q}]$ of sufficiently large depth $q = r^* + |V(G)| +1 $. Since $\vec{s}$ and $\vec{t}$ have the same stable coloring under \pluskwl, the trees $L_1$ and $L_2$ are isomorphic (by \cref{enc}). Let $\theta$ be an isomorphism from $L_1$ to $L_2$. 
	
	We prove the following equivalent statement. If $L_1$ and $L_2$ are isomorphic, then for all $i \geq 0$, $\deltaunr[G,\vec{s},i] = \deltaunr[G,\vec{t},i]$.
	The proof is by induction on $i$. The base case $i = 0$ follows trivially by comparing the isomorphism types of $\vec{s}$ and $\vec{t}$. 
	
	For the inductive case, let $j \in [k]$. Let $\vec{X}_j$ be the set of $j$-neighbors of $\vec{s}$. Similarily, let $\vec{Y}_j$ be the set of $j$-neighbors of $\vec{t}$. Our goal is to construct, for every $j \in [k]$, a corresponding bijection $\sigma_j$ between $\vec{X}_j$ and $\vec{Y}_j$ satisfying the following conditions.
	
	\begin{enumerate}
		\item For all $\vec{x}$ in $\vec{X}_j$, $\vec{x}$ is a local $j$-neighbor of $\vec{s}$ if and only if $\sigma_j(\vec{x})$ is a local $j$-neighbor of $\vec{t}$. 
		\item For all $\vec{x}$ in $\vec{X}_j$, $\deltaunr[G,\vec{x},i-1] = \deltaunr[G,\sigma_j(\vec{x}),i-1]$, i.e., $\vec{x}$ and $\sigma_j(\vec{x})$ are identically colored after $i-1$ rounds of \deltakwl.
	\end{enumerate} 
	
	From the definition of $\deltaunr$ trees, the existence of such $\sigma_1,\dots,\sigma_k$ immediately implies the desired claim $\deltaunr[G,\vec{s},i] = \deltaunr[G,\vec{t},i]$. First, we show the following claim. 
	
	\begin{claim}
		Let $\vec{C}$ be a color class in the stable coloring of $G$ under \pluskwl. 
		Let $j \in [k]$. Then, $|\vec{C} \cap \vec{X}_j| = |\vec{C} \cap \vec{Y}_j|$.
	\end{claim}
	
	\begin{proof}
		Either $|\vec{C} \cap \vec{X}_j| = |\vec{C} \cap \vec{Y}_j|= 0$, in which case we are done. Otherwise, assume without loss of generality that $|\vec{C} \cap \vec{X}_j| \neq 0$. Let $\vec{x}$ in $\vec{C} \cap \vec{X}_j$. Since $G$ is connected, we can start from the root $s$ of $L_1$, go down along $j$-labeled edges, and reach a vertex $x$ such that $x$ corresponds to the tuple $\vec{x}$. Let $w$ be the parent of $x$, and let $\vec{w}$ be the tuple corresponding to $w$. Note that $\vec{x}$ is a local $j$-neighbor of $\vec{w}$. Moreover, the depth of $\vec{w}$ is at most $n-1$. Hence, the height of the subtree of $L_1$ rooted at $w$ is at least $q - (n-1) > r^*$.  
		
		Consider the tuple $\vec{z}$ corresponding to the vertex $z = \theta(w)$ in $L_2$. Observe that the path from the root $t$ of $L_2$ to the vertex $z =\theta(w)$ consists of $j$-labeled edges. Therefore, $\vec{z}$ is $j$-neighbor of $\vec{t}$, and hence $\vec{z}$ in $\vec{Y}_j$. The stable colorings of $\vec{w}$ and $\vec{z}$ under \pluskwl are identical, because the subtrees rooted at $w$ and $z$ are of depth more than $r^*$. Let $\vec{C}$ denote the common color class of $\vec{w}$ and $\vec{z}$, in the stable coloring of $G$ under \pluskwl.
		
		Since $\vec{x}$ is a local neighbor of $\vec{w}$, the agreement of the $\#$ function values ensures that the number of $j$-neighbors (local or global) of $\vec{w}$ in $\vec{C}$ is equal to the number of $j$-neighbors (local or global) of $\vec{z}$ in $\vec{C}$. Finally, the set of $j$-neighbors of $\vec{w}$ is equal to the set of $j$-neighbors of $\vec{s}$, which is $\vec{X}_j$. Similarily, the set of $j$-neighbors of $\vec{z}$ is equal to the set of $j$-neighbors of $\vec{t}$, which is $\vec{Y}_j$. 
		Hence, $|\vec{C} \cap \vec{X}_j| = |\vec{C} \cap \vec{Y}_j|$. 
	\end{proof} 
	
	Moreover, for each $j \in [k]$, the number of local $j$-neighbors of $\vec{s}$ in $\vec{C} \cap \vec{X}_j$ is equal to the number of local $j$-neighbors of $\vec{t}$ in $\vec{C} \cap \vec{Y}_j$. Otherwise, we could perform one more round of \pluskwl and derive different colors for $\vec{s}$ and $\vec{t}$, a contradiction. 
	
	Hence, we can devise the required bijection $\sigma_j = \sigma_j^L \,\dot\cup\, \sigma_j^G$ as follows. We pick an arbitrary bijection $\sigma_j^L$ between the set of local $j$-neighbors of $\vec{s}$ inside $\vec{C}$ and the set of local $j$-neighbors of $\vec{t}$ inside $\vec{C}$. We also pick an arbitrary bijection $\sigma_j^G$ between the set of global $j$-neighbors of $\vec{s}$ inside $\vec{C}$ and the set of global $j$-neighbors of $\vec{t}$ inside $\vec{C}$. Clearly, $\sigma_j$ satisfies the first stipulated condition. By induction hypothesis, the second condition is also satisifed. Hence, we can obtain a desired bijection $\sigma_j$ satisfying the two stipulated conditions. Since we obtain the desired bijections $\sigma_1,\dots,\sigma_k$, this finishes the proof of the lemma. 
\end{proof}

Finally, since for a graph $G=(V,E)$,  $M^{\delta,\ndelta}_i(\vec{v}) = M^{\delta,\ndelta}_i(\vec{w})$ implies $M^{\delta,+}_i(\vec{v}) = M^{\delta,+}_i(\vec{w})$  for all $\vec{v}$ and $\vec{w}$ in $V(G)^k$ and $i\geq 0$, it holds that $\deltakwlm \sqsubseteq \pluskwlm$. Together with~\cref{pluseqtodelta} above, this finishes the proof of~\cref{app:loco}.

\section{Details on experiments and additional results}\label{app:exp}

Here we give details on the experimental study of~\cref{exp}.

\subsection{Datasets, graph kernels, and neural architectures}\label{datasets}

\begin{table}[h]
	\begin{center}
		\caption{Dataset statistics and properties, $^\dagger$---Continuous vertex labels following~\cite{Gil+2017}, the last three components encode 3D coordinates.}
		\resizebox{1.0\textwidth}{!}{ 	\renewcommand{\arraystretch}{1.05}
			\begin{tabular}{@{}lcccccc@{}}\toprule
				\multirow{3}{*}{\vspace*{4pt}\textbf{Dataset}}&\multicolumn{6}{c}{\textbf{Properties}}\\
				\cmidrule{2-7}
				& Number of  graphs & Number of classes/targets & $\varnothing$ Number of vertices & $\varnothing$ Number of edges & Vertex labels & Edge labels \\ \midrule
				$\textsc{Enzymes}$       & 600               & 6                 & 32.6                             & 62.1                          & \cmark  & \xmark           \\
				$\textsc{IMDB-Binary}$   & 1\,000              & 2                 & 19.8                             & 96.5                          & \xmark   & \xmark          \\
				$\textsc{IMDB-Multi}$         & 1\,500               & 3                & 13.0                             & 65.9                          & \xmark    & \xmark         \\
				$\textsc{NCI1}$          & 4\,110              & 2                 & 29.9                             & 32.3                          & \cmark   & \xmark          \\
				$\textsc{NCI109}$        & 4\,127              & 2                 & 29.7                             & 32.1                          & \cmark    & \xmark         \\
				$\textsc{PTC\_FM}$       & 349               & 2                 & 14.1                             & 14.5                          & \cmark    & \xmark         \\
				$\textsc{Proteins}$      & 1\,113              & 2                 & 39.1                             & 72.8                          & \cmark    & \xmark         \\
				$\textsc{Reddit-Binary}$ & 2\,000              & 2                 & 429.6                            & 497.8                         & \xmark     & \xmark        \\ \midrule
				
				$\textsc{Yeast}$       &   79\,601&	2&	21.5&	22.8                & \cmark  & \cmark           \\
				$\textsc{YeastH}$       &  79\,601&	2&	39.4&	40.7                  & \cmark  & \cmark           \\
				$\textsc{UACC257}$       &      39\,988&	2&	26.1&	28.1                      & \cmark  & \cmark           \\
				$\textsc{UACC257H}$       &  39\,988&	2&	46.7&	48.7      & \cmark  & \cmark           \\
				$\textsc{OVCAR-8}$       &      	40\,516&	2&	26.1&	28.1             & \cmark  & \cmark           \\
				$\textsc{OVCAR-8H}$       &        40\,516	&2&	46.7&	48.7              & \cmark  & \cmark           \\
				\midrule        
				$\textsc{Zinc}$       &  249\,456 &12	&23.1 &	24.9     & \cmark  & \cmark           \\
				$\textsc{Alchemy}$       & 202\,579 &12	& 10.1 &	10.4 	     & \cmark  & \cmark           \\
				$\textsc{QM9}$       &129\,433  &12	& 18.0 &	18.6     & \cmark (13+3D)$^\dagger$  & \cmark (4)          \\
				\bottomrule
		\end{tabular}}
		\label{ds}
	\end{center}
\end{table}

In the following, we give an overview of employed datasets, (baselines) kernels, and (baseline) neural architectures.

\begin{description}
	\item[Datasets] To evaluate kernels, we use the following, well-known, small-scale \textsc{Enzymes}~\cite{Sch+2004,Bor+2005}, \textsc{IMDB-Binary}, \textsc{IMDB-Multi}~\cite{Yan+2015a}, \textsc{NCI1}, \textsc{NCI109}~\cite{Wal+2008}, \textsc{PTC\_FM}~\cite{Hel+2001}\footnote{\url{https://www.predictive-toxicology.org/ptc/}}, \textsc{Proteins}~\cite{Dob+2003,Bor+2005}, and \textsc{Reddit-Binary}~\cite{Yan+2015a} datasets. To show that our kernels also scale to larger datasets, we additionally used the mid-scale \textsc{Yeast}, \textsc{YeastH}, \textsc{UACC257}, \textsc{UACC257H}, \textsc{OVCAR-8}, \textsc{OVCAR-8H}~\cite{Yan+2008}\footnote{\url{https://sites.cs.ucsb.edu/~xyan/dataset.htm}} datasets. For the neural architectures we used the large-scale molecular regression datasets \textsc{Zinc}~\cite{Dwi+2020,Jin+2018a} and \textsc{Alchemy}~\cite{Che+2020}.
	We opted for not using the 3D-coordinates of the \textsc{Alchemy} dataset to solely show the benefits of the (sparse) higher-order structures concerning graph structure and discrete labels.
	To further compare to the (hierarchical) \kgnn~\cite{Mor+2019} and \kign~\cite{Mar+2019}, and show the benefits of our architecture in presence of continuous features, we used the \textsc{QM9}~\cite{Ram+2014,Wu+2018} regression dataset.\footnote{We opted for comparing on the \textsc{QM9} dataset to ensure a fair comparison concerning hyperparameter selection.}
	To study data efficiency, we also used smaller subsets of the \textsc{Zinc} and \textsc{Alchemy} dataset. That is, for the \textsc{Zinc 10k} (\textsc{Zink 50k}) dataset, following~\cite{Dwi+2020}, we sampled 10\,000 (50\,000) graphs from the training, and 1\,000 (5\,000) from the training and validation split, respectively. For \textsc{Zinc 10k}, we used the same splits as provided by~\cite{Dwi+2020}. For the \textsc{Alchemy 10k} (\textsc{Alchemy 50k}) dataset, as there is no fixed split available for the full dataset\footnote{Note that the full dataset is different from the contest dataset, e.g., it does not provide normalized targets, see \url{https://alchemy.tencent.com/}.}, we sampled the (disjoint) training, validation, and test splits uniformly and at random from the full dataset. See~\cref{ds} for dataset statistics and properties.\footnote{All datasets can be obtained from~\url{http://www.graphlearning.io}.} 
	
	\item[Kernels] We implemented the \localkwl, \pluskwl, \deltakwl, and  \kwl kernel for $k$ in $\{2,3\}$. We compare our kernels to the Weisfeiler-Leman subtree kernel (\wl)~\cite{She+2011}, the Weisfeiler-Leman Optimal Assignment kernel (\wloa)~\cite{Kri+2016}, the graphlet kernel~\cite{She+2009} (\gr), and the shortest-path kernel~\cite{Bor+2005} (\shp). All kernels were (re-)implemented in \CC[11]. For the graphlet kernel we counted (labeled) connected subgraphs of size three. 
	
	\item[Neural architectures] We used the \gin and \gineps architecture~\cite{Xu+2018b} as neural baselines. For data with (continuous) edge features, we used a $2$-layer MLP to map them to the same number of components as the node features and combined them using summation (\gine and \gineeps). For the evaluation of the neural architectures of~\cref{neural}, \localkwln, \deltakwln, \kwln, we implemented them using \textsc{PyTorch Geometric}~\cite{Fey+2019}, using a  Python-wrapped \CC[11] preprocessing routine to compute the computational graphs for the higher-order GNNs. We used the \gineps layer to express $f^{W_1}_{\text{mrg}}$ and $f^{W_2}_{\text{aggr}}$ of~\cref{gnngeneral}. Finally, we used the \textsc{PyTorch}~\cite{Pas+2019} implementations of the $3$-\textsf{IGN}~\cite{Mar+2019}, and $1$-$2$-\gnn, $1$-$3$-\gnn, $1$-$2$-$3$-\gnn~\cite{Mor+2019} made available by the respective authors. 
	
	For the \textsc{QM9} dataset, we additionally used the \mpnn architecture as a baseline, closely following the setup of~\cite{Gil+2017}. For the \gineeps and the \mpnn architecture, following~\citeauthor{Gil+2017}~\cite{Gil+2017}, we used a complete graph, computed pairwise $\ell_2$ distances based on the 3D-coordinates, and concatenated them to the edge features. We note here that our intent is not the beat state-of-the-art, physical knowledge-incorporating architectures, e.g., \textsf{DimeNet}~\cite{Kli+2020} or \textsf{Cormorant}~\cite{And+2019}, but to solely show the benefits of the (local) higher-order architectures compared to the corresponding ($1$-dimensional) GNN. For the $\delta$-$2$-\textsf{GNN}, to implement~\cref{encode}, for each $2$-tuple we concatenated the (two) node and edge features, computed pairwise $\ell_2$ distances based on the 3D-coordinates, and a one-hot encoding of the (labeled) isomorphism type. Finally, we used a $2$-layer MLP to learn a joint, initial vectorial representation.
\end{description}
The source code of all methods and evaluation procedures is available at \url{https://www.github.com/chrsmrrs/sparsewl}.

\begin{table}[t]\centering		
	\caption{Classification accuracies in percent and standard deviations on medium-scale datasets.}
	\label{t2l}	
	\resizebox{0.75\textwidth}{!}{ 	\renewcommand{\arraystretch}{1.05}
		\begin{tabular}{@{}c <{\enspace}@{}lcccccc@{}}	\toprule
			
			& \multirow{3}{*}{\vspace*{4pt}\textbf{Method}}&\multicolumn{6}{c}{\textbf{Dataset}}\\\cmidrule{3-8}
			& & {\textsc{Yeast}}         &  {\textsc{YeastH}}      & {\textsc{UACC257}}           & {\textsc{UACC257H}}       & {\textsc{OVCAR-8}}           & {\textsc{OVCAR-8H}}      \\	\toprule
			& \textsf{$1$-WL}            &  88.8 \scriptsize $< 0.1$       & 88.8 \scriptsize $< 0.1$ & 96.8 \scriptsize $< 0.1$ & 96.9  \scriptsize $< 0.1$ & 96.1 \scriptsize $< 0.1$ & 96.2 \scriptsize $< 0.1$ \\
			\cmidrule{2-8}	
			\multirow{2}{*}{\rotatebox{90}{Neural}}   & \textsf{GINE}  & 88.3  \scriptsize $< 0.1 $ & 88.3 \scriptsize $< 0.1$  & 95.9 \scriptsize $< 0.1$  &  95.9  \scriptsize $< 0.1$  & 94.9   \scriptsize $< 0.1$  & 94.9 \scriptsize $< 0.1 $ \\
			& \textsf{GINE-$\varepsilon$}  &  88.3 \scriptsize $< 0.1$ & 88.3  \scriptsize $< 0.1$   & 95.9   \scriptsize $< 0.1$  & 95.9 \scriptsize $< 0.1$  & 94.9
			  \scriptsize $< 0.1$ & 94.9 \scriptsize $< 0.1 $ \\
			\cmidrule{2-8}	
			\multirow{2}{*}{\rotatebox{90}{Local}}   & \textsf{$\delta$-$2$-LWL}  & 89.2 \scriptsize $< 0.1$ & 88.9 \scriptsize $< 0.1$ &  97.0 \scriptsize $< 0.1$ & 96.9 \scriptsize $< 0.1$ & 96.4 \scriptsize $< 0.1$ & 96.3 \scriptsize $< 0.1$ \\
			& \textsf{$\delta$-$2$-LWL$^+$}  & \textbf{95.0} \scriptsize $< 0.1$ & \textbf{95.7}  \scriptsize $< 0.1$  & \textbf{97.4} \scriptsize $< 0.1$ & \textbf{98.1} \scriptsize $< 0.1$ & \textbf{ 97.4} 
			\scriptsize $< 0.1$ &\textbf{97.7}  \scriptsize $< 0.1$ \\
			\bottomrule
	\end{tabular}}
\end{table}

\begin{table}[t]\centering	\renewcommand{\arraystretch}{1.1}
	\caption{Training versus test accuracy of local and global kernels.}
	\label{t3}
	\resizebox{1.0\textwidth}{!}{ 	\renewcommand{\arraystretch}{1.05}
		\begin{tabular}{@{}c <{\enspace}@{}lcccccccc@{}}	\toprule
			
			& \multirow{3}{*}{\vspace*{4pt}\textbf{Set}}&\multicolumn{8}{c}{\textbf{Dataset}}\\\cmidrule{3-10}
			& & {\textsc{Enzymes}}         &  {\textsc{IMDB-Binary}}      & {\textsc{IMDB-Multi}}           & {\textsc{NCI1}}       & {\textsc{NCI109}}           & 
			{\textsc{PTC\_FM}}         & {\textsc{Proteins}}         &
			{\textsc{Reddit-Binary}  } 
			\\	\toprule
			\multirow{2}{*}{\rotatebox{25}{\textsf{$\delta$-2-WL}}}    		
			& Train & 91.2  & 83.8  &57.6 & 91.5   & 92.4  &  74.1 & 85.4        & --  \\ 
			
			& Test & 37.5   &  68.1 &   47.9    &      67.0   &  67.2  &   61.9  & 75.0 & \textsc{--} \\              
			\cmidrule{2-10}
			\multirow{2}{*}{\rotatebox{25}{\textsf{$\delta$-$2$-LWL}}}    		
			& Train  & 98.8  &  83.5 & 59.9  & 98.6 & 99.1 & 84.0  &  84.5 &  92.0  \\ 
			& Test  &  56.6  &  73.3  & 50.2 & 84.7 &   84.2  & 60.3  &   75.1 & 89.7  \\   
			\cmidrule{2-10}
			\multirow{2}{*}{\rotatebox{25}{\textsf{$\delta$-$2$-LWL$^+$}}}    		
			& Train  & 99.5  & 95.1 &  86.5  & 95.8 &94.4  & 96.1  & 90.9 &  96.2  \\ 
			& Test   	&  52.9 &  75.7 &  62.5  &   91.4   &  89.3     &62.6  &   79.3 &   91.1    \\       
			\bottomrule
	\end{tabular}}
\end{table}

\begin{table}[t]\centering	
	\caption{Mean MAE (mean std. MAE, logMAE) on large-scale (multi-target) molecular regression tasks.}	\label{t2n_app}	
	\resizebox{.95\textwidth}{!}{ 	\renewcommand{\arraystretch}{1.05}
		\begin{tabular}{@{}c <{\enspace}@{}lcccccc@{}}	\toprule
			
			& \multirow{3}{*}{\vspace*{4pt}\textbf{Method}}&\multicolumn{6}{c}{\textbf{Dataset}}\\\cmidrule{3-8}
			&  & {\textsc{Zinc} (10k)}         &  {\textsc{Zinc} (50k)}   &  {\textsc{Zinc (Full)}}    & {\textsc{alchemy (10k)}}     & {\textsc{alchemy (50k)}}   & {\textsc{alchemy (Full)}}       \\	\toprule
			\multirow{4}{*}{\rotatebox{90}{Baseline}}
			& \gineeps  & \textbf{0.278} \scriptsize $\pm 0.022$    & 0.145 \scriptsize $\pm  0.006  $ & 0.084                                                                                                                                                                                                                                     
			\scriptsize $\pm 0.004 $  & 0.185 {\scriptsize $\pm 0.007$} -1.864  \scriptsize $\pm 0.062$ 	& 0.127
			{\scriptsize $\pm 0.004 $} -2.415 {\scriptsize $\pm 0.053$}
			& 0.103 {\scriptsize $\pm 0.001$} -2.956 {\scriptsize $\pm 0.029$} \\	
			\cmidrule{2-8}
			&  \textsf{$2$-WL-GNN}      &  0.399  \scriptsize $\pm 0.006$
			& 0.357  \scriptsize $\pm 0.017$  & 0.133 \scriptsize $\pm 0.013$   &  0.149  {\scriptsize $\pm 0.004$}  -2.609  {\scriptsize $\pm  0.029$}
			& 0.105   {\scriptsize $\pm 0.001$}  -3.139  \scriptsize $\pm 0.020 $  & {0.093                                                                                                                                                                                                                                       
				\scriptsize $\pm 0.001$}  {-3.394                                                                                                                                                                                                                                    
				\scriptsize $\pm 0.035$}                                                                                                                                \\
			&  $\delta$-$2$-\textsf{GNN}      & 0.374 \scriptsize $\pm 0.022$
			& 0.150 \scriptsize $\pm 0.064 $ & \textbf{0.042} {\scriptsize $\pm 0.003$}   & \textbf{0.118}  {\scriptsize $\pm 0.001 $} -2.679 {\scriptsize $\pm   0.044$}  & \textbf{0.085} {\scriptsize $\pm 0.001$} -3.239 {\scriptsize $\pm 0.023$} &\textbf{0.080}                                                                                                                                                                                                                                                                                                                                                                                                                                                                                  {\scriptsize $\pm 0.001$} -3.516 {\scriptsize $\pm 0.021$} \\

			\cmidrule{2-8}	
			\multirow{2}{*}{\rotatebox{90}{}}   & $\delta$-$2$-\textsf{LGNN} &  0.306 \scriptsize $\pm0.044$
			& \textbf{0.100} \scriptsize $\pm 0.005$  & 0.045  \scriptsize $\pm 0.006$ & 0.122 {\scriptsize $\pm 0.003$} -2.573 {\scriptsize $\pm 0.078$}  & 0.090 {\scriptsize $\pm 0.001$} -3.176 {\scriptsize $\pm 0.020$} & 0.083 {\scriptsize $\pm 0.001$} -3.476  {\scriptsize $\pm 0.025$} \\
			\bottomrule
o	\end{tabular}}
\end{table}

\begin{table}[t]\centering	\renewcommand{\arraystretch}{1.05}
	\caption{Overall computation times for the whole datasets in seconds  (Number of iterations for \textsf{$1$-WL}, \textsf{$2$-WL}, \textsf{$3$-WL}, \textsf{$\delta$-$2$-WL}, \textsf{WLOA}, \textsf{$\delta$-$3$-WL}, \textsf{$\delta$-$2$-LWL}, and \textsf{$\delta$-$3$-LWL}: 5), \textsc{Oot}--- Computation did not finish within one day (24h), \textsc{Oom}--- Out of memory.}
	\label{t1_app}
	\resizebox{1.0\textwidth}{!}{ 	\renewcommand{\arraystretch}{1.1}
		\begin{tabular}{@{}c <{\enspace}@{}lcccccccc@{}}	\toprule
			& \multirow{3}{*}{\vspace*{4pt}\textbf{Graph Kernel}}&\multicolumn{8}{c}{\textbf{Dataset}}\\\cmidrule{3-10}
			& & {\textsc{Enzymes}}         &  {\textsc{IMDB-Binary}}      & {\textsc{IMDB-Multi}}           & {\textsc{NCI1}}       & {\textsc{NCI109}}           & 
			{\textsc{PTC\_FM}}         & {\textsc{Proteins}}         &
			{\textsc{Reddit-Binary}  } 
			\\	\toprule
			\multirow{4}{*}{\rotatebox{90}{\hspace*{-3pt}Baseline}} & \gr      &<1 & <1&<1 &1 &1 &<1 &<1 & 2 \\ 
			& \shp & <1 & <1 & <1 & 2 & 2 & <1 & <1 & 1\,035 \\
			& \textsf{$1$-WL}          & <1 & <1 & <1 & 2 & 2 & <1 & <1 & 2 \\
			& \textsf{WLOA}          & <1 & <1 & <1 & 14 & 14 & <1 & 1 & 15 \\
			\cmidrule{2-10}	
			\multirow{4}{*}{\rotatebox{90}{Global}} 	&
			\textsf{$2$-WL}        & 302 & 89 & 44 & 1\,422 & 1\,445& 11 &  14\,755 & \textsc{Oom} \\
			& \textsf{$3$-WL}        	& 74\,712 & 18\,180 & 5\,346 & \textsc{Oot}&\textsc{Oot} & 5\,346&\textsc{Oom} &\textsc{Oom} \\
			\cmidrule{2-10}
			
			& \textsf{$\delta$-$2$-WL}  	& 294& 89& 44 & 1\,469& 1\,459& 11 &14\,620 & \textsc{Oom}  \\
			
			& \textsf{$\delta$-$3$-WL}          		&64\,486 & 17\,464 & 5\,321 & \textsc{Oot}&\textsc{Oot} & 1119 &\textsc{Oom} &\textsc{Oom} \\                                                   
			
			\cmidrule{2-10}		
			\multirow{4}{*}{\rotatebox{90}{Local}}    		
			
			& \textsf{$\delta$-$2$-LWL}         	&29 & 25& 20 & 101 & 102 & 1 & 240 & 59\,378 \\      
			& \textsf{$\delta$-$2$-LWL$^+$}         	& 35 & 31 & 24 & 132& 132 & 1 & 285 &84\,044\\       
			
			& \textsf{$\delta$-$3$-LWL} 	&4\,453 &3\,496 &2\,127 & 18\,035 & 17\,848 & 98 &\textsc{Oom} &\textsc{Oom} \\ 
			& \textsf{$\delta$-$3$-LWL$^+$}         	&4\,973 & 3\,748 & 2\,275 &20\,644 & 20\,410& 105 & \textsc{Oom}  &\textsc{Oom} \\    
			\bottomrule
	\end{tabular}}
\end{table}

\begin{table}[t]\centering	\renewcommand{\arraystretch}{1.1}
	\caption{Overall computation times for the whole datasets in seconds on medium-scale datasets (Number of iterations for \textsf{$1$-WL}, \textsf{$\delta$-$2$-LWL}, and \textsf{$\delta$-$3$-LWL}: 2).}
	\label{t1l_app}
	\resizebox{.75\textwidth}{!}{ 	\renewcommand{\arraystretch}{1.05}
		\begin{tabular}{@{}c <{\enspace}@{}lcccccc@{}}	\toprule
			
			& \multirow{2}{*}{\vspace*{4pt}\textbf{Graph Kernel}}&\multicolumn{6}{c}{\textbf{Dataset}}\\\cmidrule{3-8}
			& & {\textsc{Yeast}}         &  {\textsc{YeastH}}      & {\textsc{UACC257}}           & {\textsc{UACC257H}}       & {\textsc{OVCAR-8}}           & {\textsc{OVCAR-8H}}      \\	\toprule
			
			& \textsf{$1$-WL}            & 11   & 19  & 6 & 10& 6 & 10 \\
			
			\cmidrule{2-8}	
			\multirow{2}{*}{\rotatebox{90}{Local}}   & \textsf{$\delta$-$2$-LWL}  &1\,499  & 5\,934  & 1\,024 & 3\,875	 & 1\,033  & 4\,029 \\
			& \textsf{$\delta$-$2$-LWL$^+$}         	& 2\,627 &7\,563 & 1\,299 & 4\,676& 1\,344 & 4\,895 \\    
			\bottomrule
	\end{tabular}}
\end{table}

\subsection{Experimental protocol and model configuration}\label{protocol}

In the following, we describe the experimental protocol and hyperparameter setup.
\begin{description}
	\item[Kernels] For the smaller datasets (first third of~\cref{ds}), for each kernel, we computed the (cosine) normalized gram matrix. We computed the classification accuracies using the $C$-SVM implementation of \textsc{LibSVM}~\cite{Cha+11}, using 10-fold cross-validation. 
	We repeated each 10-fold cross-validation ten times with different random folds, and report average accuracies and standard deviations. For the larger datasets (second third of~\cref{ds}), we computed explicit feature vectors for each graph and used the linear $C$-SVM implementation of \textsc{LibLinear}~\cite{Fan+2008}, again using 10-fold cross-validation (repeated ten times). 
	Following the evaluation method proposed in~\cite{Mor+2020}, in the both cases, the $C$-parameter was selected from $\{10^{-3}, 10^{-2}, \dotsc, 10^{2},$ $10^{3}\}$ using a validation set sampled uniformly at random from the training fold (using 10\% of the training fold). Similarly, the number of iterations of the \wl, \wloa, \localkwl, \pluskwl, and \kwl were selected from $\{0,\dotsc,5\}$ using the validation set. Moreover, for the \pluskwl, we only added the additional label function $\#$ on the last iteration to prevent overfitting.
	We report computation times for the \wl, \wloa, \localkwl, \pluskwl, and \kwl with five refinement steps. All kernel experiments were conducted on a workstation with an \text{Intel Xeon E5-2690v4} with 2.60\si GHz and 384\si GB of RAM running \text{Ubuntu 16.04.6 LTS} using a single core. Moreover, we used the GNU \CC Compiler 5.5.0 with the flag \texttt{--O2}. 
	\item[Neural architectures] For comparing to kernel approaches, see \cref{t2,t2l}, we used 10-fold cross-validation, and again used the approach outlined in~\cite{Mor+2020}. The number of components of the (hidden) node features in $\{ 32, 64, 128 \}$ and the number of layers in $\{ 1,2,3,4,5\}$ of the \gin (\gine) and \gineps (\gineeps) layer were again selected using a validation set sampled uniformly at random from the training fold (using 10\% of the training fold). We used mean pooling to pool the learned node embeddings to a graph embedding and used a $2$-layer MLP for the final classification, using a dropout layer with $p = 0.5$ after the first layer
	of the MLP. We repeated each 10-fold cross-validation ten times with different random folds, and report the average accuracies and standard deviations. Due to the different training methods, we do not provide computation times for the GNN baselines. 
	
	For the larger molecular regression tasks, \textsc{Zinc} and \textsc{Alchemy}, see~\cref{t2n_app}, we closely followed the hyperparameters found in~\cite{Dwi+2020} and~\cite{Che+2020}, respectively, for the \gineeps layers. That is, for \textsc{Zinc}, we used four \gineeps layers with a hidden dimension of 256 followed by batch norm and a $4$-layer MLP for the joint regression of the twelve targets, after applying mean pooling. For \textsc{Alchemy} and \textsc{QM9}, we used six layers with 64 (hidden) node features and a set2seq layer~\cite{Vin+2016} for graph-level pooling, followed by a $2$-layer MLP for the joint regression of the twelve targets. We used exactly the same hyperparameters for the (local) $\delta$-$2$-\textsf{LGNN}, and the dense variants $\delta$-$2$-\textsf{GNN} and $2$-\textsf{WL-GNN}. 
	
	For \textsc{Zinc}, we used the given train, validation split, test split, and report the MAE over the test set. For the \textsc{Alchemy} and \textsc{Qm9} datasets, we uniformly and at random sampled 80\% of the graphs for training, and 10\% for validation and testing, respectively. Moreover, following~\cite{Che+2020,Gil+2017}, we normalized the targets of the training split to zero mean and unit variance. We used a single model to predict all targets. Following~\cite[Appendix C]{Kli+2020}, we report mean standardized MAE and mean standardized logMAE. We repeated each experiment five times (with different random splits in case of \textsc{Alchemy} and \textsc{Qm9}) and report average scores and standard deviations. 
	
	To compare training and testing times between the $\delta$-$2$-\textsf{LGNN}, the dense variants the $\delta$-$2$-\textsf{GNN} and $2$-\textsf{WL-GNN}, and the (1-dimensional) \gineeps layer, we trained all four models on \textsc{Zinc (10k)} and \textsc{Alchemy (10k)} to convergence, divided by the number of epochs, and calculated the ratio with regard to the average epoch computation time of the $\delta$-$2$-\textsf{LGNN} (average computation time of dense or baseline layer divided by average computation time of the $\delta$-$2$-\textsf{LGNN}). All neural experiments were conducted on a workstation with four Nvidia Tesla V100 GPU cards with 32GB of GPU memory running Oracle Linux Server 7.7.
\end{description}

\end{document}